\documentclass[12pt]{article}
\usepackage{amscd,amssymb,amsmath,latexsym,enumerate}
\usepackage[mathscr]{euscript}
\usepackage{epsfig}
\usepackage{fancybox}
\usepackage{verbatim}

\usepackage{color}
\newcommand{\red}{\textcolor[rgb]{0.90,0.00,0.00}} 
\title{The non-commutative topology \\
of two-dimensional dirty superconductors}

\author{Giuseppe De Nittis$^1$, Hermann Schulz-Baldes$^{2,3}$
\\
\\
{\small $^1$ Departamento de Matem\'atica, Instituto de F\'isica, Pontificia Universidad Cat\'olica,
Santiago, Chile}
\\
{\small $^2$ Department Mathematik, Friedrich-Alexander-Universit\"at Erlangen-N\"urnberg, Germany}
\\
{\small $^3$ Instituto de Matem\'aticas, UNAM, Unidad Cuernavaca, Mexico}
}

\date{ }

\newtheorem{theo}{Theorem}
\newtheorem{defini}{Definition}
\newtheorem{proposi}{Proposition}
\newtheorem{lemma}{Lemma}

\newcommand{\CM}{{\mathbb C}}
\newcommand{\NM}{{\mathbb N}}
\newcommand{\RM}{{\mathbb R}}

\newcommand{\TM}{{\mathbb T}}
\newcommand{\ZM}{{\mathbb Z}}
\newcommand{\PM}{{\mathbb P}}

\newcommand{\UM}{{\mathbb U}}

\newcommand{\Aa}{{\cal A}}
\newcommand{\Pp}{{\cal P}}

\newcommand{\EE}{{\bf E}}
\newcommand{\Bb}{{\cal B}}

\newcommand{\Ee}{{\cal E}}
\newcommand{\Ff}{{\cal F}}
\newcommand{\Gg}{{\cal G}}

\newcommand{\Uu}{{\cal U}}
\newcommand{\Vv}{{\cal V}}

\newcommand{\Oo}{{\cal O}}
\newcommand{\Tr}{\mbox{\rm Tr}}
\newcommand{\Tt}{{\cal T}}

\newcommand{\Jj}{{\cal J}}

\newcommand{\Ll}{{\cal L}}

\newcommand{\Hh}{{\cal H}}

\newcommand{\one}{{\bf 1}}

\newcommand{\TR}{{\rm Tr}} 
\newcommand{\Ch}{{\rm Ch}} 
\newcommand{\Ind}{{\rm Ind}} 
\newcommand{\diag}{{\rm diag}}

\newcommand{\redu}{{\mbox{\rm\tiny red}}}
\newcommand{\ph}{{\mbox{\rm\tiny ph}}}

\newcommand{\frakc}{{\mathfrak c}}
\newcommand{\frakd}{{\mathfrak d}}

\begin{document}

\maketitle

\begin{abstract}

Non-commutative analysis tools have successfully been applied to the integer quantum Hall effect, in particular for a proof of the stability of the Hall conductance in an Anderson localization regime and of the bulk-boundary correspondence. In this work, these techniques are implemented to study two-dimensional dirty superconductors described by Bogoliubov-de Gennes Hamiltonians. After a thorough presentation of the basic framework and the topological invariants, Kubo formulas for the thermal, thermoelectric and spin Hall conductance are analyzed together with the corresponding edge currents.

\vspace{.1cm}


\end{abstract}



\section{Introduction}
\label{sec-overview}

Topological insulators have been the object of intense experimental and theoretical investigations over the last decade, and more recently also in the mathematical physics community. Many of the analytical tools developed for the study of integer quantum Hall systems could be adapted and extended to these novel topological systems. Several theoretical elements have become common themes of the field, in particular the identification of the topological invariants, their link to non-dissipative response coefficients and the bulk-boundary correspondence. For recent reviews and a vast literature, we refer to \cite{PS,Sch2}. In our opinion, a thorough mathematical treatment of physical phenomena in topological superconductors is lacking to date and this work aims to partially fill this gap in dimension two. 

\vspace{.2cm}

Just as a large part of the physics literature, we focus on quadratic fermionic Hamiltonians described by an effective Bogoliubov-de Gennes (BdG) Hamiltonian on a particle-hole Hilbert space. These operators typically do not have particle and charge conservation due to the presence of a non-trivial pairing potential. Numerous such potentials are of physical interest and are reviewed in Section~\ref{sec-BdGgen}. Some of them have further symmetries, like a U$(1)$  or SU$(2)$ rotational invariance in the spin degrees of freedom. Such symmetries can be reduced out in a way that is independent of the dimension of physical space. While this symmetry analysis is somewhat standard by now, {\it e.g.} \cite{BLS,AZ,HHZ}, it is nevertheless included in Section~\ref{sec-BdGgen} for sake of completeness. In Section~\ref{sec-alg} it is then combined with the C$^*$-algebraic framework of \cite{Bel,BES,PS} for the description of homogeneous media. Up to this point, the treatment is independent of the dimension of physical space, but starting from Section~\ref{sec-invariants} we decided to restrict to two-dimensional tight-binding systems for which the main topological invariant is then the Chern number of the BdG Fermi projection. There is no principle difficulty in transposing also the numerous (strong and weak) invariants as discussed in \cite{SRFL,PS} to BdG Hamiltonians in other dimensions (that is actually already covered by \cite{GS}), but carrying this out would have made this paper too encyclopedic and we hope that the most important concepts are communicated on the example of two-dimensional systems. On the other hand, we do go beyond earlier papers because a description of disordered systems is entirely covered. Merely an Anderson localization condition as in \cite{BES,PS} has to be satisfied for the existence of the BdG Chern numbers.

\vspace{.2cm}

The second part of the paper, that is Sections~\ref{sec-BBC} to \ref{sec-TEQHE},  then deals with various physical effects linked to a non-trivial topological invariant. In fact, it is shown how the bulk invariant (Chern number) determines the non-dissipative Hall response coefficient for the mass transport, the charge and spin transport (provided charge and spin are conserved), as well as the low temperature behavior of the thermal Hall transport. The latter so-called thermal quantum Hall effect \cite{SFi} is probably the most interesting case, as no further symmetry has to be imposed on the BdG Hamiltonian (which is hence in the Cartan-Altland-Zirnbauer (CAZ) Class D). Unfortunately, we were unable to give a rigorous derivation of the Kubo formula for the thermal Hall conductance and hence merely analyze the (presumably correct) formula from \cite{SSt,QSN,SF}. The spin quantum Hall effect \cite{SMF} is, on the other hand, merely a superposition of several (conventional) quantum Hall effects in each of the eigenspaces of the spin operator. The bulk-boundary correspondence (BBC) for the mass transport is a direct consequence of earlier results \cite{KRS,PS}. The BBC for the charge and spin transport is then a special case under supplementary symmetry constraints. Based on the physical interpretation of the boundary currents in these cases, we then derive a formula for the thermal boundary currents which, by the BBC, is then again quantized with a coefficient given by the bulk invariant. In our opinion, this considerably clarifies previous treatments \cite{SFi,Vis,Kit0}. As a final comment, let us mention that another physical effect in Class D are the Majorana zero modes attached to half-flux vortices \cite{RG}. This is not dealt with here, but in our previous paper \cite{DS}. 

\vspace{.2cm}

\noindent {\bf Acknowledgements:}  The first draft of this manuscript dates back about 5 years, and it was somewhat more ambitious than the present manuscript. While this may say something about the authors, it is also due to the fact that physical concepts are harder to unfold and lay out than mathematical results. Along the way we profited from discussions with numerous colleagues, in particular, Martin Zirnbauer and Emil Prodan, whom we wish to thank. Moreover, we thank Varghese Mathai and Guo Thiang for inviting us both to an interesting conference in Adelaide and to contribute to the proceedings.  G.D.'s research is supported by the  grant \emph{Iniciaci\'{o}n en Investigaci\'{o}n 2015} - $\text{N}^{\text{o}}$ 11150143 funded  by FONDECYT, and that of H.S.-B. by the DFG.

\section{Generalities on BdG Hamiltonians}
\label{sec-BdGgen}

\subsection{BdG Hamiltonian in tight-binding representation}
\label{sec-tightbindingBdGgen}

Let us begin by presenting the two-dimensional tight-binding BdG Hamiltonian which provides an effective description of electrons (quasi-particles) in a superconductor. Each single particle will be described by a one-particle Hamiltonian $h$ acting on the one-particle Hilbert space $\Hh=\ell^2(\ZM^2)\otimes\CM^L$. Depending on what is to be described, the fiber $\CM^L$ may contain a factor $\CM^{2s+1}$ to describe the spin $s\in \NM/2$ of the particle, and a fiber $\CM^2$ to model a bipartite lattice structure like for the hexagon lattice, as well as any further local internal degrees of freedom of the particle, like simply several participating orbitals at every site. For sake of simplicity, we only consider the case $\CM^L=\CM^{2s+1}$ here. Of course, the case of spin $\frac{1}{2}$ is physically most relevant. In bra-ket notation, $h$ is of the form
\begin{equation}
\label{eq-kineticfirstquant}
h\;=\;\sum_{n,n'\in\ZM^2}\,\sum_{l,l'=1}^L\;h_{l,l'}(n,n')\,|n,l\rangle\langle n',l'|
\;=\;
\sum_{n,n'\in\ZM^2}\,|n\rangle \,h(n,n')\,\langle n'|
\;.
\end{equation}
Here $h_{l,l'}(n,n')$ are complex numbers such that $h=h^*$ and $h(n,n')=(h_{l,l'}(n,n'))_{l,l'=1,\ldots,L}$ is an $L\times L$ matrix. Furthermore, $|n\rangle$ is the partial isometry to (spin) states at $n\in\ZM^2$. The first main assumption on $h$ will be that it is of finite range $R$, namely $h(n,n')=0$ for $|n-n'|>R$. From a mathematical point of view, this locality condition could be somewhat relaxed, but this is irrelevant for the physics. Later on, the second main assumption is that $h$ is space homogeneous (see Section~\ref{sec-alg} for details). This does allow $h$ to contain a random potential, for example.

\vspace{.2cm}

Next let us (canonically) second quantize $h$ to an operator ${\bf h}$ on the fermionic Fock space $\Ff=\Ff(\Hh)$ associated to $\Hh=\ell^2(\ZM^2)\otimes\CM^L$. If the creation and annihilation operators in the state $|n,l\rangle$ are denoted by $\frakc^*_{n,l}$ and $\frakc_{n,l}$, then the second quantization of \eqref{eq-kineticfirstquant} leads to
$$
{\bf h}\;=\;\sum_{n,n'\in\ZM^2}\,\sum_{l,l'=1}^L\;h_{l,l'}(n,n')\,\frakc^*_{n,l} \frakc_{n',l'}
\;=\;
\sum_{n,n'\in\ZM^2}\,\frakc_n^* \,h(n,n')\,\frakc_{n'}
\;=\;
\frakc^*\,h\,\frakc
\;.
$$
Here, in the second formula $\frakc_n$ denotes a vector (spinor) $\frakc_n=(\frakc_{n,l})_{l=1,\ldots,L}$ of annihilation operators, while in the third formula $\frakc$ is the vector $\frakc=(\frakc_n)_{n\in\ZM^2}$. There is no summability associated to this vector and it is merely used in order to have the compact notation ${\bf h}=\frakc^*h\frakc$ for the summation given above. The second quantized operator ${\bf h}$ clearly commutes with the number operator ${\bf N}=\frakc^*\frakc$ so that it conserves the particle number. Furthermore, let us point out that the anti-commutation relations for the $\frakc$'s imply, with $h^T$ denoting the transpose of $h$, 
$$
\frakc^*\,h\,\frakc
\;=\;
-\,\frakc\,h^T\,\frakc^*\,+\,\Tr(h)\,\one_\Ff
\;=\;
-\,\frakc\,\overline{h}\,\frakc^*\,+\,\Tr(h)\,\one_\Ff
\;,
$$
as long as $\Tr(h)$ is finite ({\it e.g.} for a finite sublattice of $\ZM^2$). The term $\Tr(h)\,\one_\Ff$ is merely a constant shift in energy which will be neglected in the spirit of renormalization, even if $\Tr(h)$ is infinite as it gives no contribution to commutators. This is equivalent to working with a second quantized 
\begin{equation}
\label{eq-kineticsecondquant}
{\bf h}
\;=\;
\frac{1}{2}\;\frakc^*\,h\,\frakc\;-\;\frac{1}{2}\,\frakc\,\overline{h}\,\frakc^*
\;.
\end{equation}

\vspace{.2cm}

In BCS theory there is now an (attractive) interaction between the particles which can be written as a quartic term in the $\frakc$'s. A typical interaction of this type is the term $\sum_n \frakc^*_n\frakc_n\frakc_n^*\frakc_n$ in the Hubbard model. Such a term leads to a full many body problem. In the Bogoliubov version of Hartree-Fock theory \cite{dG,BLS} this term is replaced by a quadratic term which models the creation and annihilation of Cooper pairs, and is calculated self-consistently \cite{dG}. This approximation produces a \emph{quadratic Hamiltonian} ${\bf H}$ which, by virtue of the self-consistency equations, has the particle-hole symmetric form
\begin{equation}
\label{eq-HBdG}
{\bf H}
\;=\;
\frac{1}{2}\;\frakc^*\,h\,\frakc\;-\;\frac{1}{2}\,\frakc\,\overline{h}\,\frakc^*
\;+\;
\frac{1}{2}\;\frakc^*\,\Delta\,\frakc^*
\;-\;
\frac{1}{2}\;\frakc\,\overline{\Delta}\,\frakc
\;.
\end{equation}
Here again $\Delta=(\Delta(n,n'))_{n,n'\in\ZM^2}$ is given by $L\times L$ matrices $\Delta(n,n')$ which satisfy the finite range condition $\Delta(n,n')=0$ for $|n-n'|>R$, and $\frakc^*\Delta \frakc^*$ is a short notation for
$$
\frakc^*\,\Delta\,\frakc^*
\;=\;
\sum_{n,n'\in\ZM^2}\,\frakc_n^* \,\Delta(n,n')\,\frakc^*_{n'}
\;=\;
\sum_{n,n'\in\ZM^2}\,\sum_{l,l'=1}^L\;\Delta_{l,l'}(n,n')\,\frakc^*_{n,l} \frakc^*_{n',l'}
\;.
$$
Furthermore $\overline{\Delta}$ in \eqref{eq-HBdG} denotes complex conjugation. The Hamiltonian \eqref{eq-HBdG} is the starting point of our analysis. Let us point out that the self-adjointness of ${\bf H}$ combined with the particular particle-hole symmetric form \eqref{eq-HBdG} imposes 
\begin{equation}
\label{eq-DeltaClassD}
\Delta^*
\;=\;
-\,\overline{\Delta}
\;.
\end{equation}
The factor $\frac{1}{2}$ in the third and fourth summand of \eqref{eq-HBdG} is an artefact introduced for later notational convenience.

\subsection{Particle-hole symmetry of BdG Hamiltonians}
\label{sec-BdGfirstquant}

The Hamiltonian \eqref{eq-HBdG} is quadratic in the creation and annihilation operators and can thus be treated by a suitable first quantized Hamiltonian. Indeed,  \eqref{eq-HBdG} can be rewritten in a matrix form  as
\begin{equation}
\label{eq-BdGlink}
{\bf H}
\;=\;
\frac{1}{2}\,
\begin{pmatrix}
\frakc^* & \frakc
\end{pmatrix}\begin{pmatrix}
h & \Delta \\ -\overline{\Delta} & -\overline{h}
\end{pmatrix}
\begin{pmatrix}
\frakc \\ \frakc^*
\end{pmatrix}
\;,
\end{equation}
where the appearing matrix is an operator on the particle-hole Hilbert space $\Hh_{\mbox{\tiny\rm ph}}=\Hh\otimes\CM^2_{\mbox{\tiny\rm ph}}$ and the operator-valued scalar product is also understood on this Hilbert space. The factor $\CM^2_{\mbox{\tiny\rm ph}}$ is called the particle-hole fiber. For any chemical potential $\mu\in\RM$, one then has
\begin{equation}
\label{eq-BdGfirst}
{\bf H}\,-\,\mu\,{\bf N}
\;=\;
\frac{1}{2}\,
\begin{pmatrix}
\frakc^* & \frakc
\end{pmatrix}
H(\mu)
\begin{pmatrix}
\frakc \\ \frakc^*
\end{pmatrix}
\;,
\qquad
H(\mu)
\;=\;
\begin{pmatrix}
h-\mu & \Delta \\ -\overline{\Delta} & -\overline{h}+\mu
\end{pmatrix}
\end{equation}
The operator matrix $H(\mu)$ is the BdG Hamiltonian on $\Hh_{\mbox{\tiny\rm ph}}$ (first quantized, but with particle-hole fiber).  For operators given by $2\times 2$ block matrices such as the BdG Hamiltonian \eqref{eq-BdGfirst}, one speaks of a representation in the particle-hole grading. In order to keep notations light, we just write ${\bf H}$ for ${\bf H}-\mu{\bf N}$, $H$ for $H(\mu)$ and even $h$ for $h-\mu$. This simply means that we assume $\mu=0$ from now on, unless the dependence on $\mu$ is analyzed. The BdG Hamiltonian $H$ satisfies an even particle-hole symmetry (PHS):
\begin{equation}
\label{eq-BdGsymmetry}
K^*\,\overline{H}\,K
\;=\;
-\,H
\;,
\qquad
K\;=\;
\begin{pmatrix} 0 & \one \\ \one &  0 \end{pmatrix}
\;.
\end{equation}
This implies that for even and odd real functions $f_\pm:\RM\to\RM$ satisfying $f_\pm(-x)=\pm f_\pm(x)$ one has
$$
K^*\,f_\pm (\overline{H})\,K\;=\;\pm\,f_\pm(H)\;.
$$
Decomposing the Fermi-Dirac function $f_\beta(E)=(1+e^{-\beta E})^{-1}$ into even and odd parts, this implies, for example,
$$
K^*\,f_\beta (\overline{H})\,K\;=\;\one\,-\,f_\beta(H)\;.
$$

\subsection{Examples of pairing potential}
\label{sec-interaction}

The operator $\Delta$ on $\Hh$ is called the \emph{pairing  potential} because it dictates the creation and annihilation of Cooper pairs in \eqref{eq-HBdG}. It is sometimes also called the pairing field.  Often the pairing potential is chosen to be translation invariant and the numbers characterizing $\Delta$ are then called the superconducting order parameters. In this section, some examples of such translation invariant pair potentials are presented for two-dimensional systems. The list below does not cover all cases studied in the literature \cite{WSS,Sca}, but hopefully the most important ones. Generally, the one-particle Hamiltonian is simply the discrete Laplacian $h=V_1+V_1^*+V_2+V_2^*$ where $V_1$ and $V_2$ are the shift operators  on $\ell^2(\ZM^2)$ (without magnetic fields and naturally extended to $\Hh=\ell^2(\ZM^2)\otimes\CM^L$).  Using $s^1,s^2,s^3$ for a spin $s$  representation, the list of pairing potentials is then: 
\begin{align}
& \Delta_s\;=\;
\delta_s\one\,\otimes\;\imath s^2\;,
&&  s=\tfrac{1}{2}\;\;\mbox{(singlet $s$-wave)}
\\
& \Delta_{s^*}\;=\;
\delta_{s^*}\,(V_1+V_1^*+V_2+V_2^*)\,\otimes\, \imath\, s^2\;,
&&  s=\tfrac{1}{2}\;\;\mbox{(singlet extended $s$-wave)}
\\
& \Delta_{p_x}\;=\;
\delta_{p_x}\,(V_1-V_1^*)\,\otimes\,s^1\;,
&&  s=\tfrac{1}{2}\;\;\mbox{(spinful $p_x$-wave)}
\label{eq-px}
\\
& \Delta_{p\pm \imath p}\;=\;
\delta_{p}\,(V_1-V_1^*\pm\imath(V_2-V_2^*) )\;,
&&  s=0\;\;\mbox{(spinless $p\pm\imath p$-wave)}
\label{eq-pip}
\\
& \Delta_{p}\;=\;\Delta_{p\pm\imath p}\,\otimes\,s^1\;,
&&  s=\tfrac{1}{2}\;\;\mbox{(spinful $p\pm\imath p$-wave)}
\\
& \Delta'_{p}\;=\;\delta_{p'}\,((V_1-V_1^*)\,\otimes\,\one_L\pm\imath(V_2-V_2^*)\,\otimes\,s^3 )
\;,
&&  s=\tfrac{1}{2}\;\;\mbox{(triplet $p\pm\imath p$-wave)}
\\
& \Delta_{d_{xy}}\;=\;
\delta_{d_{xy}}\,(V_1-V_1^*)(V_2-V_2^*)\,\otimes\,\imath\, s^2 \;,
&&  s=\tfrac{1}{2}\;\;\mbox{(singlet $d_{xy}$-wave)}
\\
& \Delta_{d_{x^2-y^2}}\;=\;
\delta_{d_{x^2-y^2}}\,(V_1+V_1^* -V_2-V_2^*)\,\otimes\,\imath\,s^2 \;,
&&  s=\tfrac{1}{2}\;\;\mbox{(singlet $d_{x^2-y^2}$-wave)}
\\
& \Delta_{d\pm \imath d}\;=\;\Delta_{d_{x^2-y^2}}\pm\imath \,\Delta_{d_{xy}}
 \;,
&&  s=\tfrac{1}{2}\;\;\mbox{(singlet $d\pm\imath d$-wave)}
\label{eq-did}
\;.
\end{align}
All the constants $\delta$ are real so that one readily checks that \eqref{eq-DeltaClassD} holds in all cases. Here $p$-wave pair potentials correspond to hopping terms which are anti-symmetric under the change $V_j\leftrightarrow V_j^*$, while $s$-wave and $d$-wave pair potentials are symmetric under this change. Furthermore, the $s$-wave is rotation symmetric and the $d$-wave odd under a 90 degree rotation $(V_1,V_2,V_1^*,V_2^*)\leftrightarrow (V_2,V_1^*,V_2^*,V_1)$. These symmetries result from the atomic orbitals relevant for the description of a given material. A graphic representation is given in Fig.~1.


\begin{figure}
\begin{center}
\includegraphics[height=2.85cm]{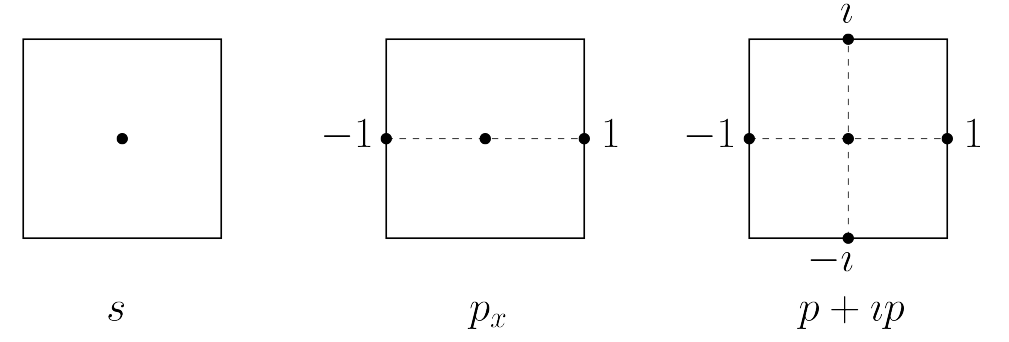}
\hspace{-.1cm}
\includegraphics[height=2.9cm]{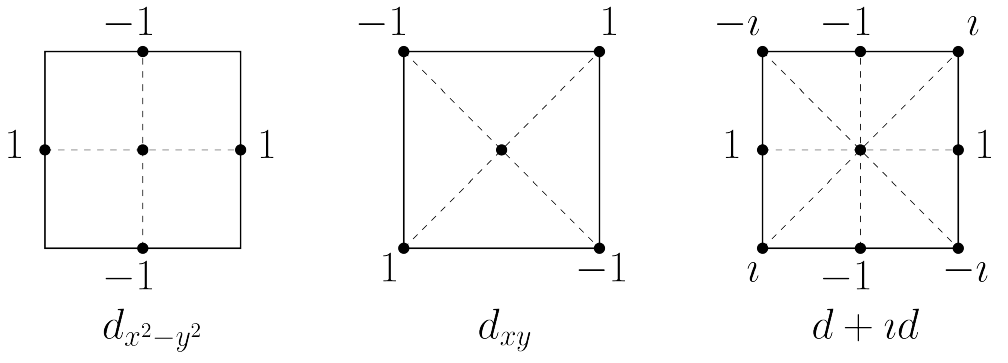}
\caption{\it Graphic representation of various pair potentials.}
\end{center}
\end{figure}

\subsection{Canonical and Bogoliubov transformations}
\label{sec-Bogo}

In this section, we show how to diagonalize the quadratic Hamiltonians by canonical transformations by following standard textbook treatments, {\it e.g.} \cite{BR}. The PHS \eqref{eq-BdGsymmetry} of the Hamiltonian \eqref{eq-BdGfirst} can be interpreted as follows: $\imath\,H$ is in the Lie algebra of the group
$$
\Gg\;=\;\left\{A\in \mbox{\rm GL}(\Hh_{\mbox{\tiny\rm ph}})\, : \, K^* \overline{A}K=A\right\}
\;.
$$
Let $\Uu=\Gg\cap\UM(\Hh_{\mbox{\tiny\rm ph}})$ denote the unitaries in this group: 
$$
\Uu\;=\;\left\{W\in \mbox{\rm GL}(\Hh_{\mbox{\tiny\rm ph}})\, : \,  W^*=W^{-1}\;,\;\;
K^*\overline{W}K=W\right\}
\;.
$$
Again $\Uu$ is a group.  Explicitly, one has
\begin{equation}
\label{eq-Bogogroup}
\Uu\;=\;
\left\{
W=
\left(\begin{array}{cc}
 u & v \\
 \overline{v} & \overline{u}
 \end{array}\right)
\;:
\;
u^*u+v^T\overline{v}=\one\;,\;\;u^*v+v^T\overline{u}=0
\right\}
\;.
\end{equation}
Now, given $W\in\Uu$, one can define
\begin{equation}
\label{eq-Bogtrafo}
\left(\begin{array}{c}
 \frakd \\
 \frakd^* \end{array}\right)
 \;=\;
 W
 \left(\begin{array}{c}
 \frakc \\
 \frakc^* \end{array}\right)
 \;=\;
 \left(\begin{array}{c}
 u\, \frakc+v\, \frakc^* \\
 \overline{v}\, \frakc+\overline{u}\, \frakc^* 
 \end{array}\right)
 \;, 
\end{equation}
where in the last equality $W$ is given as in the representation \eqref{eq-Bogogroup}. The particular form of $W$ assures that $\frakd$ and $\frakd^*$ are indeed mutually adjoint and that the CAR relations for $\frakd$ and $\frakd^*$ hold due to the condition $W^*=W^{-1}$, which in turn is equivalent to the one given in \eqref{eq-Bogogroup}. A standard question is now whether \eqref{eq-Bogtrafo} can be implemented by a unitary opertor ${\bf U}_W$ on Fock space in the sense that $\frakd={\bf U}_W^*\, \frakc\, {\bf U}_W$ (attention: ${\bf U}_W$ is not quadratic in $\frakc$; it depends on $W$ but it is not associated to $W$ via a formula like \eqref{eq-BdGfirst}). For a finite system, this is always possible, but in infinite dimension one has to impose a condition. It is sufficient to require $v$ to be Hilbert-Schmidt (Shale's theorem) \cite{Sh,BLS}. Then the unitary ${\bf U}_W$ is called a Bogoliubov transformation, while $W$ is usually called the associated canonical transformation. Hence $\Uu$ is also called the group of canonical transformations. 

\vspace{.2cm}

For the remainder of the section let us suppose that $\dim(\Hh)=N$ is finite. Then one readily checks that 
$$
\Vv\,\Uu\,\Vv^*\;=\;
\mbox{\rm O}(2N)
\;,
\qquad
\Vv\;=\;
\sqrt{\frac{\imath}{2}}
\begin{pmatrix}
\one & -\imath\,\one \\
\imath\,\one & -\one
\end{pmatrix}
\;
$$
by means of relations $\Vv^*=\Vv^{-1}$ and $\Vv^T\Vv=-\Vv \Vv^T=K$.
Furthermore, $H$ can be diagonalized by a canonical transformation $W$:
$$
W\, H\, W^*
\;=\;
\left(\begin{array}{cc}
 D& 0\\
   0 & -D
 \end{array}\right)
\;,
$$
where $D$ is diagonal and real valued. Using this particular canonical transformation, one has
\begin{equation}
\label{eq-Hdiag}
\begin{aligned}
{\bf H}
\;&=\;
\frac{1}{2}\;
\left(\begin{array}{cc}
\frakd^*  &
\frakd 
\end{array}\right)
\left(\begin{array}{cc}
 D& 0\\
   0 & -D
 \end{array}\right)
\left(\begin{array}{c}
\frakd  \\
\frakd^* 
\end{array}\right)
\\ \;&=\;
\frac{1}{2}\;
{\bf U}_W^\ast\left[\left(\begin{array}{cc}
\frakc^* &
\frakc
\end{array}\right)
\left(\begin{array}{cc}
 D& 0\\
   0 & -D
 \end{array}\right)
\left(\begin{array}{c}
\frakc  \\
\frakc^* 
\end{array}\right)\right] {\bf U}_W
\;.
\end{aligned}
\end{equation}
This will be used for the calculation of the thermal equilibrium state in the next section.

\subsection{Gibbs state of BdG Hamiltonians}

Also in this section, $\dim(\Hh)=N$ is finite. Recall that the Gibbs state $\omega_\beta$ of ${\bf H}$ at a given inverse temperature $\beta$ (and chemical potential $\mu=0$ already absorbed in ${\bf H}$) is defined by
$$
\omega_{\beta}( {\bf A})
\;=\;
\frac{1}{Z_\beta}\; \Tr_{\Ff}\left( {\bf A}\, e^{-\beta  {\bf H}}\right)
\;,
\qquad
Z_\beta\;=\;\Tr_{\Ff}\left(e^{-\beta  {\bf H}}\right)
\;,
$$
where ${\bf A}$ is an operator on $\Ff$. Because ${\bf H}$ is quadratic in the creation and annihilation operators, it is well-known that $\omega_\beta$ is quasi-free (Wick's theorem applies, {\it e.g.} \cite{BLS}). In particular, $\omega_\beta$ is completely specified by the associated one-particle density matrix $\Gamma_\beta$ which is the operator on $\Hh_{\mbox{\tiny\rm ph}}=\Hh\otimes\CM^2_{\mbox{\tiny\rm ph}}$ defined by the two-point functions: 
$$
\Gamma_\beta
\;=\;
\omega_\beta
\left(\begin{array}{cc}
 \frakc\frakc^* & \frakc\frakc \\
\frakc^* \frakc^* & \frakc^* \frakc
 \end{array}\right)
\;=\;
\sum_{n,n'\in\ZM^2}
\,|n\rangle\,
\begin{pmatrix}
 \omega_\beta(\frakc_{n}\frakc_{n'}^*) & \omega_\beta(\frakc_{n}\frakc_{n'}) \\ \omega_\beta(\frakc^*_{n} \frakc^*_{n'}) & \omega_\beta(\frakc_{n}^* \frakc_{n'})
\end{pmatrix}
\,\langle n'|
 \;.
 $$
This is a positive operator on $\Hh_{\mbox{\tiny\rm ph}}$ which satisfies
$$
K\, \overline{\Gamma_\beta}\, K
\;=\;
{\bf 1}-\Gamma_\beta
\;.
$$
Furthermore, for any canonical transformation $W$ one has
$$
W \,\Gamma_\beta\, W^*
\;=\;
\omega_{\beta}
\left(\begin{array}{cc}
 \frakd\frakd^*& \frakd\frakd\\
   \frakd^* \frakd^* & \frakd^* \frakd
 \end{array}\right)
 \;.
$$
Using this for the $W$ diagonalizing $H$ so that \eqref{eq-Hdiag} holds, one deduces the standard formula for the quasi-free Gibbs states of an ideal Fermi gas: 
$$
W \,\Gamma_\beta\, W^*
\;=\;
\begin{pmatrix}
f_\beta(D) & 0 \\ 0 & f_\beta(-D)
\end{pmatrix}
\;,
\qquad
f_\beta(E)\;=\;
(e^{\beta E}+1)^{-1}
\;.
$$
After some algebra this implies that $\Gamma_\beta$ is given by the Fermi-Dirac function of $H$:
$$
\Gamma_\beta
\;=\;
f_\beta(H)
\;.
$$
Therefore, if ${\bf A}$ is an observable that is quadratic in $\frakc$ and $\frakc^*$ so that there exists a one-particle observable $A$ on $\Hh_{\mbox{\tiny\rm ph}}$ (using the same formula as \eqref{eq-BdGlink}), then
\begin{equation}
\label{eq-Fac1/2}
\omega_{\beta}( {\bf A})
\;=\;
\frac{1}{2}\;\Tr_{\Hh_{\mbox{\tiny\rm ph}}}(f_\beta(H)\,A)
\;.
\end{equation}
Further below, this formula allows to go to the thermodynamic limit by introducing the trace per unit volume whenever $H$ and $A$ are space homogeneous. 

\subsection{BdG Hamiltonians with charge conservation}
\label{sec-Charge}

Particle conservation of the second quantized Hamiltonian is $[{\bf H},{\bf N}]=0$. For the BdG Hamiltonian this becomes charge conservation 
$$
[H,Q]\;=\;0
\;,
$$
where the charge operator on $\Hh_\ph$ is
$$
Q
\;=\;
\begin{pmatrix}
\one & 0 \\ 0 & -\one
\end{pmatrix}
\;.
$$
This means that the anti-particles in the BdG representation have the opposite charge from the particle, which for sake of concreteness is chosen to be positive. Clearly charge conservation is equivalent to a vanishing pair creation potential and thus $H=\diag(h-\mu,-\overline{h}+\mu)$ and the passage to the BdG formalism results merely in a doubling.

\subsection{$\mbox{\rm SU}(2)$ invariant BdG Hamiltonians}
\label{sec-SU2inv}

Let us recall that the spin of the particle is denoted by $s\in\NM/2$ and that $L=2s+1$. Associated to $s$ is an irreducible representation of the Lie algebra $\mbox{\rm su}(2)$ on $\CM^L$. Let $s^1,s^2,s^3$ denote the hermitian, traceless $L\times L$ matrices representing the 3 components of the spin operator corresponding to a basis of the Lie algebra $\mbox{\rm su}(2)$ such that the commutation relations $[s^1,s^2]=\imath\,s^3$, $[s^2,s^3]=\imath\,s^1$ and $[s^3,s^1]=\imath\,s^2$ hold. As usual, we choose the representation  with $s^1$ and $s^3$ real, and  $s^2$  purely imaginary. In Appendix~\ref{app-su2} the concrete (standard) representations used here are listed. The extension $\one\otimes s^j$ to $\Hh=\ell^2(\ZM^2)\otimes \CM^L$ will also simply be denoted by $s^j$. Then their second quantized is denoted by boldface ${\bf s}^j$, which is explicitly given by
\begin{equation}
\label{eq-spinsymbol}
{\bf s}^j
\;=\;
\sum_{n\in\ZM^2}\;\sum_{l,l'=1}^L(s^j)_{l,l'}\,\frakc^*_{n,l}\,\frakc_{n,l'}
\;=\;
\frac{1}{2}\;
\begin{pmatrix} \frakc & \frakc^* \end{pmatrix}
S^j
\begin{pmatrix} \frakc \\ \frakc^* \end{pmatrix}
\;,
\end{equation}
where
\begin{equation}
\label{eq-spinsymbol2}
S^j
\;=\;
\begin{pmatrix} s^j & 0 \\ 0 & -\,(s^j)^T \end{pmatrix}
\;.
\end{equation}
Let us observe that the second equality in \eqref{eq-spinsymbol} is exact in view of the fact that the spin matrices are traceless. These second quantizations again satisfy the $\mbox{\rm su}(2)$ relations $[{\bf s}^1,{\bf s}^2]=\imath\,{\bf s}^3$, $[{\bf s}^2,{\bf s}^3]=\imath\,{\bf s}^1$ and $[{\bf s}^3,{\bf s}^1]=\imath\,{\bf s}^2$ (this needs an algebraic check because ${\bf s}^i{\bf s}^j$ contains quartic operators in the creation and annihilation operators, which cancel in the commutator). Therefore they provide a representation of $\mbox{\rm su}(2)$ on the Fock space $\Ff$. Now $\mbox{\rm SU}(2)$ invariance of a Hamiltonian ${\bf H}$ on Fock space means by definition that for ${\bf s}(\theta)=\sum_{j=1,2,3}\theta_j{\bf s}^j$ with arbritrary real $\theta=(\theta_1,\theta_2,\theta_3)$, one has
$$
e^{\imath\,{\bf s}(\theta)}\,{\bf H}\,e^{-\imath\,{\bf s}(\theta)}
\;=\;{\bf H}
\;,
$$
or equivalently
$$
[{\bf s}^j,{\bf H}]\;=\;0\;,
\qquad
j=1,2,3
\;.
$$
If ${\bf H}$ is a quadratic Hamiltonian with BdG Hamiltonian $H$, these latter conditions are equivalent to 
\begin{equation}
\label{eq-SU2inv}
\left[
S^j,H \right]
\;=\;0
\;,
\qquad
j=1,2,3
\;.
\end{equation}
This follows from \eqref{eq-commutatorid} in Appendix~\ref{app-commutator}.

\begin{proposi}
\label{prop-SU2inv}
Let $H$ be a {\rm BdG} Hamiltonian that satisfies the $\mbox{\rm SU}(2)$ invariance {\rm \eqref{eq-SU2inv}} with respect to a spin $s$ representation. Then there are operators  $h_\redu$ 
and $\Delta_\redu$ on $\ell^2(\ZM^2)$ such that
$$
H
\;=\;
\begin{pmatrix}
h_\redu\otimes \one_L & \Delta_\redu\otimes T
\\
-\,\overline{\Delta_\redu}\otimes T & -\overline{h_\redu}\otimes\one_L
\end{pmatrix}
\;,
$$
where $T=(T_{l,l'})_{l,l'=1,\ldots,L}$ with $T_{l,l'}=\chi(l+l'=L+1)(-1)^{l+1}$ is the 
matrix having only alternating signs on the cross-diagonal. Hence $H$ decomposes into a direct sum of copies of two unitarily equivalent building blocks 
$$
H_\redu
\;=\;
\begin{pmatrix}
h_\redu & \Delta_\redu
\\
\sigma\,\overline{\Delta_\redu} & -\overline{h_\redu}
\end{pmatrix}
\;,
\qquad
H'_\redu
\;=\;
\begin{pmatrix}
h_\redu & -\,\Delta_\redu
\\
-\,\sigma\,\overline{\Delta_\redu} & -\overline{h_\redu}
\end{pmatrix}
\;,
$$
acting on the reduced particle-hole space $\Hh_{\mbox{\rm\tiny red,ph}}=\ell^2(\ZM^2)\otimes \CM^2_{\mbox{\rm\tiny ph}}$, where $\sigma=(-1)^{L}$ is $-1$ for integer spin and $1$ for half-integer spin. For half-integer spin $H$, contains $\frac{L}{2}$ copies of each 
reduced model, while for  integer spin, there are  $\frac{L+1}{2}$ copies of $H_\redu$ and $\frac{L-1}{2}$ of $H'_\redu$. 
\end{proposi}

\noindent {\bf Proof.} Writing out the commutators explicitly, one realizes that \eqref{eq-SU2inv} is equivalent to
\begin{equation}
\label{eq-partcom}
[h,s^j]\;=\;0\;,
\qquad
\Delta (s^j)^T
\;=\;-\,s^j\Delta
\;,
\qquad
j=1,2,3
\;.
\end{equation}
Schur's lemma then implies that $h=h_\redu\otimes \one_L$ for some $h_\redu$. As $(s^1)^T=s^1$, $(s^2)^T=-s^2$ and $(s^3)^T=s^3$, the second set of equations becomes the following anti-commutation and commutation relations
$$
\{\Delta,s^1\}\;=\;\{\Delta,s^3\}\;=0\;,
\qquad [\Delta,s^2]\;=\;0\;.
$$
Now let us  consider $\Delta$ as an $L\times L$ matrix of operators on $\ell^2(\ZM^2)$ and denote its $l$th row and column by $R_l$ and $C_l$ respectively. Since $s^3$ is diagonal, the relation $\{\Delta,s^3\}=0$ reads
$$
\left(\begin{array}{c}
s\, R_1 \\
 (s-1)\, R_2 \\
\vdots \\
-s\, R_L
\end{array}\right)
\;+\;
 \left(\begin{array}{cccc}
s\, C_1 &(s-1)\, C_2 & \cdots & -s\,  C_L
 \end{array}\right)
\;=\;0
\;.
$$
This forces $\Delta$ to have non-zero entries only on the cross-diagonal, namely with entries given by operators on $\ell^2(\ZM^2)$,
\begin{equation}
\label{eq-u1invariance}
\Delta\;=\;
\left(\begin{array}{ccccc}
 &&&& \Delta_1\\
 & &&\Delta_2&\\
  &&.\cdot\, \dot{ } &&
  \\
  &\Delta_{L-1}&& &\\
  \Delta_L&&&&
 \end{array}\right)
 \;.
\end{equation}
The remaining relations  $\{\Delta,s^1\}=0=[\Delta,s^2]$ can be rewritten in terms of $s_-$ and $s_+$ defined in Appendix~\ref{app-su2}, giving
\begin{equation}
\label{eq-matr}
\Delta\, s_+\,+\,s_-\,\Delta\;=\;0\;,
\qquad 
\Delta\,s_-\,+\,s_+\,\Delta\;=\;0\;. 
\end{equation}
The first of these relations reads, with the $\alpha_l$ given in Appendix~\ref{app-su2},
$$
\left(\begin{array}{ccccc}
0&\alpha_1\, C_1& \cdots & \alpha_{L-2}\, C_{L-2} &\alpha_{L-1}\, C_{L-1}
 \end{array}\right)
 \;+\;
 \left(\begin{array}{c}
0 \\
\alpha_1\, R_1
\\
\vdots
\\
\alpha_{L-2}\, R_{L-2} \\
\alpha_{L-1}\, R_{L-1}
 \end{array}\right)
 \;=\;0
 \;.
$$
This implies $\alpha_j\, \Delta_j+\alpha_{L-j}\, \Delta_{j+1}=0$ for $j=1,\ldots,L-1$. Since $\alpha_j=\alpha_{L-j}=\sqrt{j(L-j)}$, one also concludes that $\Delta_{j+1}=-\Delta_j$, so that $\Delta=\Delta_\redu \otimes T$ for some operator $\Delta_\redu$. It can then be checked that the second equation in \eqref{eq-matr} does not add any further constraint.
\hfill $\Box$

\vspace{.2cm}

Let us point out that the matrix $T$ appearing in Proposition~\ref{prop-SU2inv} is symmetric if the spin is integer ($L$ odd), and anti-symmetric if the spin is half-integer ($L$ even). Hence a comparison with \eqref{eq-DeltaClassD} shows that:
\begin{equation}\label{eq:PP-cnstraints}
\begin{aligned}
&-\Delta_\redu
\;=\;
(\Delta_\redu)^T
\;,
\qquad
s\;\;\mbox{\rm integer}\;\;(L\;\;\mbox{\rm odd})\;, \\
&+\Delta_\redu
\;=\;
(\Delta_\redu)^T
\;,
\qquad
s\;\;\mbox{\rm  half-integer}\;\;(L\;\;\mbox{\rm even})\;.\\
\end{aligned}
\end{equation}
Therefore the reduced (spinless) Hamiltonians $H_\redu$ and $H_\redu'$ have the same even PHS \eqref{eq-BdGsymmetry} as $H$ for integer spin, but in the case of half-integer spin, one has the symmetry 
\begin{equation}\label{eq:oddPHS}
I^*\,\overline{H_\redu}\,I\;=\;-\,H_\redu
\;,
\qquad
I
\;=\;
\begin{pmatrix}
0 & -\,\one \\ \one & \;\; 0
\end{pmatrix}
\;,
\end{equation}
and a similar relation for $H_\redu'$. This is referred to as an odd PHS and one then says that the operator $H_\redu$ is in the CAZ Class C \cite{AZ}. Furthermore, the unitary equivalence claimed above is explicitly given by
$$
J^*\,{H_\redu}\,J\;=\;H'_\redu
\;,
\qquad
J
\;=\;
\begin{pmatrix}
\one & 0 \\ 0 & -\one 
\end{pmatrix}
\;.
$$
Finally let us point out that, by the same argument as in the case without symmetries, the Fermi-Dirac function of the reduced BdG satisfies
$$
I^*\,f_\beta(\overline{H_\redu})\,I\;=\;\one-f_\beta(H_\redu)
\;.
$$

\vspace{.2cm}

As to examples of  $\mbox{\rm SU}(2)$ invariant models, let us consider the case of a spin $s=\frac{1}{2}$. Then $T$ in Proposition~\ref{prop-SU2inv} is given by $T=2\imath\,s^2$. Consequently any pair potential $\Delta$ of the form $\Delta=p(V_1,V_2)\,\imath\,s^2$ with a polynomial $p(V_1,V_2)$ that is a symmetric operator, is $\mbox{\rm SU}(2)$ invariant. In particular, in the list of Seciton~\ref{sec-interaction} the  pair potentials $\Delta_s$, $\Delta_{s^*}$, $\Delta_{d_{xy}}$, $\Delta_{d_{x^2-y^2}}$ and $\Delta_{d\pm\imath d}$ lead to $\mbox{\rm SU}(2)$ invariant BdG Hamiltonians if also $h$ is SU$(2)$ invariant.

\subsection{$\mbox{\rm U}(1)$ invariant BdG Hamiltonians}
\label{sec-U1inv}

In some models, there is not a full $\mbox{\rm SU}(2)$ rotational invariance, but only an invariance under spin rotations around one axis. Let us choose coordinates such that the $3$-axis is the rotational axis. In the case of a spin $\frac{1}{2}$ the following result was already pointed out in \cite{SRFL}.

\begin{proposi}
\label{prop-U1inv}
Let ${\bf H}$ be a BdG Hamiltonian with total spin $s$ that satisfies $[{\bf s}^3,{\bf H}]=0$.  Then the first quantized Hamiltonian $H$ decomposes into a direct sum of $L$ reduced Hamiltonians on $\ell^2(\ZM^2)\otimes \CM^2_{\mbox{\rm\tiny ph}}$ which have no symmetry at all {\rm (}Class A{\rm )}.
\end{proposi}

\noindent {\bf Proof.} Other than in the proof of Proposition~\ref{prop-SU2inv}, only the relation $j=3$ of  \eqref{eq-partcom} holds. The first commutation relation $[s^3,h]=0$ implies that $h=\mbox{diag}(h_1,\ldots,h_L)$, while $\{s^3,\Delta\}=0$ implies again \eqref{eq-u1invariance}. As $h$ is diagonal and $\Delta$ is anti-diagonal, one readily deduces the claim. For each $l=1,\ldots,L$, the reduced Hamiltonian has the form $H^{(l)}_{\redu}=\binom{h_l\;\;\Delta_l}{\Delta_l^*\;-\overline{h}_l}$ with $h_l$ and $\Delta_l$  operators on $\ell^2(\ZM^2)$. Since the potentials $\Delta_l$ are not subjected to additional constraints, the  reduced Hamiltonians $H_l$ 
do not inherit any symmetry properties,  in general. 
\hfill $\Box$

\vspace{.2cm}

Examples of a $\mbox{\rm U}(1)$ invariant BdG Hamiltonian are given by spin independent $h$ and a pair potential $\Delta=p(V_1,V_2)\,s^1$ with a polynomial $p(V_1,V_2)$ that is an anti-symmetric operator. Indeed, $s^1$ anti-commutes with $s^3$. The pair potential $\Delta_p$ given in Seciton~\ref{sec-interaction} is of this form.

\subsection{BdG Hamiltonians with TRS}
\label{sec-TRS}

The time reversal operator $\Theta$  on the one-particle Hilbert space $\Hh=\ell^2(\ZM^2)\otimes\CM^L$  is given by the complex conjugation $C$ (a real structure on $\Hh$ which we also simply denote by an overline $C\psi=\overline{\psi}$) and a rotation $R$ of each spin component by 180 degrees. If $C$ also acts on the spin degree of freedom, one thus has
$$
(\Theta \psi)_n\;=\;R\;\overline{\psi_n}\;,
\qquad
R\;=\;e^{\imath \pi s^2}\;.
$$
Hence $\Theta=RC=CR$ is an anti-unitary operator, namely $\Theta$ is anti-linear and satisfies $\Theta^*\Theta=\Theta\Theta^*=\one$ where $\Theta^*$ is defined by $\langle \psi|\Theta^*\phi\rangle=\langle\phi| \Theta\,\psi\rangle$. Clearly $\Theta^*=\Theta^{-1}$. Moreover, $R$ is real and one has $\Theta^2=e^{2\pi\imath s^2}C^2=(-1)^{2s}$ and $R^*=(-1)^{2s}R$ because the spectrum of $s^2$ is $\{-s,-s+1,\ldots,s\}$.  One calls the time-reversal operator even or odd depending whether $\Theta^2=1$ and $\Theta^2=-1$ respectively. Another relation of use below is that the spin transforms like angular momentum:
\begin{equation}
\label{eq-spinrev}
\Theta^{-1}\,s^j\,\Theta\;=\;-\,s^j\;,
\qquad
j=1,2,3\;.
\end{equation}

The next objective is to second quantize $\Theta$ to an anti-unitary operator ${\bf \Theta}={\bf RC}$ on fermionic Fock space $\Ff$. The complex conjugation ${\bf C}$ is just the complex conjugation on $\Ff$ inherited from $\Hh$, while ${\bf R}=\oplus_{n\geq 0} R^{\otimes n}$ is the standard second quantization of a unitary. One also has
$$
{\bf R}\;=\;e^{\imath\,\pi\,{\bf s}^2}
\;,
$$
where ${\bf s}^2$ is the second quantization of the one-particle generator $s^2$ given in \eqref{eq-spinsymbol}. Now the TRS of a second quantized Hamiltonian reads
$$
{\bf \Theta}\,{\bf H}\,{\bf \Theta}^{-1}\;=\;{\bf H}
\;.
$$
Due to the above, this can also be rewritten as
\begin{equation}
\label{eq-TRS}
{\bf R}\,{\bf H}\,{\bf R}^{-1}
\;=\;\overline{{\bf H}}
\;.
\end{equation}
Now let us restrict our attention to a BdG Hamiltonian ${\bf H}$ of the form \eqref{eq-BdGlink}. One checks ${\bf R}\,\frakc_{n,l}^*{\bf R}^{-1}=\sum_{k}R_{k,l}\frakc_{n,k}^*$. In the column/row notations of \eqref{eq-BdGlink}, this leads to ${\bf R}\,\frakc^*{\bf R}^{-1}=\frakc^*R=R^T\frakc^*$ as well as ${\bf R}\,\frakc\,{\bf R}^{-1}=R^*\frakc=  \frakc\,\overline{R}$ so that
$$
{\bf R}\,{\bf H}\,{\bf R}^{-1}
\;=\;
\frac{1}{2}\,
\begin{pmatrix}
\frakc^* & \frakc
\end{pmatrix}
\begin{pmatrix}
RhR^* & R\Delta R^T \\ -\overline{R\Delta R^T} & -\overline{RhR^*}
\end{pmatrix}
\begin{pmatrix}
\frakc \\ \frakc^*
\end{pmatrix}
\;.
$$
As $R=\overline{R}$ is real, the TRS \eqref{eq-TRS} is equivalent to $(R\oplus R)\,H\,(R\oplus R)^*=\overline{H}$ or to
\begin{equation}
\label{eq-TRS2}
R\,h\,R^{-1}\;=\;\overline{h}
\;,
\qquad
R\,\Delta\,R^{-1}\;=\;\overline{\Delta}
\;.
\end{equation}
To exploit these equations, it is best to go into the spectral representation of the unitary $R$. In the case of even TRS $R^2=\one$ so that $R=\Pi_+-\Pi_-$ where $\Pi_+$ and $\Pi_-$ are the corresponding orthogonal spectral projections of $R$. In the case of odd TRS the $R=\imath\,\Pi_+-\imath\,\Pi_-$. If the  spin is $0$, then $\Pi_-=0$ and $R=\one$. In the case of spin $\frac{1}{2}$, one has $R=2\imath s^2=\binom{0\,-1}{1\;\;0}$. The TRS can now be combined with all three cases of BdG Hamiltonians considered above, the one without symmetries (except for PHS), the one with full $\mbox{\rm SU}(2)$ invariance and the one with $\mbox{\rm U}(1)$ invariance. As the main focus of this paper is on the BdG operators without TRS (without and with $\mbox{\rm SU}(2)$ invariance), no further details will be given here.

\subsection{Resum\'e on BdG classes and associated physical phenomena}
\label{sec-classes}

In this section, let us briefly resume the classification of symmetries of BdG operators in the following table which also contains an outlook on what is to come in the remainder of the paper. We restrict to systems with half-integer spin. After the columns with symmetries appear the CAZ classes of the reduced model, as further discussed below.

\vspace{.2cm}

\label{table1}
\begin{center}
\begin{tabular}{|c|c||c|c||c||c|c|c|c|}
\hline
SU$(2)$ & U$(1)$ &PHS&TRS &  CAZ & PP & Invariant & Effect     
\\
\hline
\hline
$0$&$0$ &$+1$ &$0$ & D & $\Delta_{p\pm\imath p}$&$\ZM$ & TQHE, Maj. States  
\\
$0$&$0$ &$+1$  &$-1$&DIII &$\Delta'_p$& $ \ZM_2$  & $\ZM_2$ top.~ins. 
\\
\hline
\hline
$1$&$1$ &$-1$ &$0$ & C & $\Delta_{d\pm\imath d}$& $2\,\ZM$  &   SQHE 
\\
$1$&$1$ &$-1$ &$+1$& CI & $\Delta_{d_{xy}}$, $\Delta_{d_{x^2-y^2}}$& 
$ 0$  &   
\\
$0$&$1$ &$0$ &$0$& A & $\Delta_p$& 
$ \ZM$  &   SQHE  
\\
$0$&$1$ &$0$ &$-1$& {AIII}&  $\Delta_{p_x}$
&$0$&    
\\ \hline
\end{tabular}
\end{center}

\vspace{.2cm}

\noindent

In the remaining columns we restrict to dimension $d=2$. First follow examples of pair potentials (PP) having the symmetries of the corresponding row (we  tacitly assume that the non-interacting part $h$ of the BdG operator has the same symmetry), then the strong invariants associated with them. They can be seen as phase labels and will be discussed in detail in Section~\ref{sec-invariants}. All of this is in agreement with the standard classification table as in \cite{SRFL}, see also \cite{GS}, but the above table is not complete (even not for $d=2$). Indeed, systems with particle conservation are not listed as they are not the focus of this paper. The final column contains a list of physical effects linked to the invariants. They will be discussed in the remainder of the paper.

\vspace{.2cm}

The first two columns of the table describe the absence (with 0) or the presence (with 1) of global SU$(2)$ or U$(1)$ symmetry for the spin degrees of freedom of the full Hamiltonian \eqref{eq-BdGfirst}. When both these symmetries are absent none of the components of the spin is a conserved quantity. This implies that no reduction as described in Proposition \ref{prop-SU2inv} and Proposition \ref{prop-U1inv} is possible. In this case, the even PHS symmetry \eqref{eq-BdGsymmetry} is the only symmetry of the BdG Hamiltonian and it results merely from the rewriting of a quadratic fermionic Hamiltonian in the BdG representation. In addition, it can have a TRS. These possibilities are described in the  first two rows of the table.  When the TRS is broken,  as in the case of the first row, the model is in Class D and the strong invariant takes values in $\ZM$. The case of an odd TRS is described  in the second row. In this situation the model is in Class DIII and the related strong invariant is in $\ZM_2$. The case of an even TRS, namely of systems in Class BDI, is not listed as fermions are only subject to odd TRS (moreover, for $d=2$ Class BDI only has the trivial phase \cite{SRFL}).

\vspace{.2cm}

The last four rows of the table describe cases in which all or at least one component of the spin is conserved by the full Hamiltonian \eqref{eq-BdGfirst}. Then the reduction procedures described in Proposition \ref{prop-SU2inv} or Proposition \ref{prop-U1inv} apply.  In this situation the type of PHS and TRS along with the related CAZ class refer to  the reduced  BdG operators $H_{\redu}$ on $\Hh_{\mbox{\rm\tiny ph}}$ as given by Propositions~\ref{prop-SU2inv} and \ref{prop-U1inv}, respectively (always the fermionic half-integer spin case). More precisely, in the case of SU$(2)$ invariance (for the full Hamiltonian) described by the third and fourth rows the reduced model is subjected to an odd PHS as shown by \eqref{eq:oddPHS}. If TRS symmetry is absent the reduced operator is in Class C and the topological classification is provided by even integers $2\ZM$, see the third row of the table. In the case of pair potentials $\Delta_{d_{xy}}$ or $\Delta_{d_{x^2-y^2}}$ which are proportional to $s^2$ in the spin space, the full BdG operator commutes with the operator $R$ and so the action of the TRS just reduces to complex conjugation at level of the reduced operator. As a consequence, even though the TRS is odd for the full model due to the half-integer spin, it acts as an even TRS on the reduced operator $H_{\redu}$. This leads to the CAZ label CI that appears in the fourth row of the table above.

\vspace{.2cm}

The last two rows of the table describe the case in which only the component $s^3$ of the spin is conserved and in this case the reduced operator $H_{\redu}$ obtained in Proposition~\ref{prop-U1inv} does not acquire any PHS. In absence of TRS the reduced system is in Class A which is classified by $\ZM$, see the fifth row. For the last case let us note that the pair potential $\Delta_{p_x}$  is proportional to $s^1$ w.r.t. the spin variables of the full Hamiltonian \eqref{eq-BdGfirst} and  is therefore real under complex conjugation. If also $h$ is real, the complex part of the TRS commutes with the full BdG operator and the reduced operator $H_{\redu}$ obtained by imposing  the U$(1)$ symmetry  acquires a chiral symmetry. This leads to  the label AIII in the last row of the table above.

\vspace{.2cm}

Finally let us briefly comment on systems with charge conservation. In accordance with the discussion in Section~\ref{sec-Charge}, the pair potentials then vanish and the system is a (virtual) sum of two copies of the operator $h-\nu$ which plays the role of the reduced operator. Again, it can or cannot have a chiral or time-reversal symmetry. These cases are not listed in the table.

\section{Observable algebra for homogeneous systems}
\label{sec-alg}

In this section, we combine the BdG formalism with the algebraic description of disordered media from \cite{Bel}.

\subsection{Covariant operators}

The two magnetic translations  $u_1$ and $u_2$ on the one-particle Hilbert space $\Hh=\ell^2(\ZM^2)\otimes\CM^L$ are defined by
\begin{equation}
\label{eq-magntrans}
(u_1\psi)_n\;=\;
\; \psi_{ n-e_1}
\;,
\qquad
(u_2\psi)_n\;=\;
 e^{ \imath qB X_1 }
\; \psi_{ n-e_2}
\;,
\end{equation}
where $e_1=(1,0)$ and $e_2=(0,1)$ are the two unit vectors, $B$ is the magnetic field (orbital and not coupled to the spin) in the $3$-direction perpendicular to the system in the Landau gauge and $q$ is the particle charge, which we suppose to be positive. Because the anti-particles have negative charge $-q$, the magnetic translations naturally extend as $U_j=u_j\oplus\overline{u_j}$ to the particle-hole Hilbert space $\Hh_{\mbox{\rm\tiny ph}}=\Hh\otimes \CM^2_{\mbox{\rm\tiny ph}}$. The commutation relation of these operators is
$$
U_1\,U_2
\;=\;
\Xi
\;
U_2\,U_1
\;,
\qquad
\Xi\;=\;
\begin{pmatrix}
e^{\imath qB} & 0 \\ 0 & e^{-\imath qB}
\end{pmatrix}
\;.
$$
The main focus will be on strongly continuous families $A=(A_{\omega})_{\omega\in\Omega}$ of bounded operators on $\Hh_{\mbox{\rm\tiny ph}}$ indexed by a compact space $\Omega$ of disorder or crystaline configurations. This space is supposed to be compact and furnished with a continuous action $\tau=(\tau_1,\tau_2)$ of the translation group $\ZM^2$. The family $A$ is then supposed to satisfy the covariance relation
\begin{equation}
\label{eq-CovRel}
U_j A_{\omega}  U_j^{-1} \;=\;A_{\tau_j \omega}
\;,
\qquad j=1,2
\;.
\end{equation}
The following result shows that a covariant operators breaking charge conservation can only exists if the magnetic field takes very particular values. This is a mathematical manifestation of the Meissner effect. No magnetic fields penetrate into a supercondutor. When the external magnetic field becomes strong, a type I superconductor undergoes a phase transition to a normal metal, while a type II superconductor forms half-flux vortices which exactly compensate the exterior field. These vortices often form so-called Abrikosov lattices. After an adequate gauge transformation on the particle-hole space, one indeed obtains a model with vanishing constant magnetic field, as is well-explained in \cite{FT,VMFT}.

\begin{proposi}
\label{prop-ZeroMag}
Let $A=(A_{\omega})_{\omega\in\Omega}$ be a family of bounded operators on $\Hh_{\mbox{\rm\tiny ph}}=\ell^2(\ZM^2,\CM^{2L})$ satisfying the covariance relation \eqref{eq-CovRel}. If $[A,Q]\not=0$, then $qB$ can only take the values $0,\frac{\pi}{2},\pi,\frac{3\pi}{2}$.
\end{proposi}

\noindent {\bf Proof.} For sake of notational simplicity, let us restrict to the case where $\Omega$ consists of just one point. Let $A=\binom{a\;b}{c\;d}$ with matrix entries acting on $\ell^2(\ZM^2,\CM^{2L})$. Then write $b=\sum_{n,n'\in\ZM^2}\,|n\rangle \,b(n,n')\,\langle n'|$. The covariance relation for $b$ reads
\begin{equation}\label{eq:cov_cond}
u_j \,b\;=\;b\; \overline{u_j}\;,\qquad j=1,2
\;,
\end{equation}
which implies
$$
b(n+e_1,n'+e_1)\;=\;b(n,n')\;,
\qquad 
b(n+e_2,n'+e_2)\;=\;e^{ -\imath 2qB (n_1+n'_1)}\;b(n,n')
\;
$$
where $n_1+n'_1=(n+n')\cdot e_1$.
From these relations one infers
$$
b(n+e_1+e_2,n'+e_1+e_2)\;=\;b(n+e_2,n'+e_2)\;=\;e^{ -\imath 2qB (n_1+n'_1)}\;b(n,n')
$$
as well as
$$
b(n+e_1+e_2,n'+e_1+e_2)\,=\,
e^{ -\imath 2qB (n_1+n'_1+2)}\;b(n+e_1,n'+e_1)
\,=\,e^{ -\imath 4qB }e^{ -\imath 2qB (n_1+n'_1)}\;b(n,n')\,.
$$
If now $b(n,n')\not=0$ for some $n$ and $n'$, then indeed $e^{ -\imath 4qB }=1$.
\hfill $\Box$

\vspace{.2cm}

\noindent {\bf Remark}
When $qB=0$ (zero magnetic field) or  $qB=\pi$ (half flux), one has  $\overline{u_j}=u_j$ for both $j=1,2$ hence the conditions \eqref{eq:cov_cond} become the usual covariant relations for $u_jbu_j^*=b$ which are verified by any $b$ in the commutant $C^*(u_1,u_2)'$ of the $C^*$-algebra  generated by $u_1$ and $u_2$. This (von Neumann) algebra is non trivial. In fact, it contains the  $C^*$-algebra  generated by the shift operators $V_1$ and $V_2$ introduced in Section \ref{sec-interaction}.
\hfill $\diamond$

\vspace{.2cm}

The idea in the following is to consider $A(\omega,n)=\langle 0|A_\omega|n\rangle$ as matrix-valued symbols for covariant operator families and to construct a C$^*$-algebra out of these symbols, given by an adequate twisted crossed product \cite{Bel}. First one endows the topological vector space $C_c(\Omega\times\ZM^{2}, \mbox{\rm Mat}(2L,\CM))$ of continuous functions with compact support on $\Omega \times \ZM^{2}$ and values in $\mbox{\rm Mat}(2L,\CM)$ with a $*$-algebra structure:
\begin{equation}
\label{eq-staralg0}
AB ({\omega}, n) \;=\;
 \sum_{m\in\ZM^{2}}\,
   A(\omega, m) \,B(\tau^{-m} \omega, n-m)\,
\Xi^{n_1-m_1}
\mbox{ , }
\qquad
     A^*(\omega, n) \; = \;
      A(\tau^{-n}\omega, -n)^*
\mbox{ , }
\end{equation}
\noindent where the last $*$ denotes the adjoint matrix, $n_1-m_1=(n-m)\cdot e_1$ and $\tau^n=\tau_1^{n_1}\tau_2^{n_2}$. Here $\mbox{\rm Mat}(2L,\CM)=\mbox{\rm Mat}(L,\CM)\otimes \mbox{\rm Mat}(2,\CM)$ consists of $L\times L$ matrices acting on the spin components tensorized with $2\times 2$ matrices in the particle hole space.  For $\omega \in \Omega$, a representation of this  $*$-algebra on $\Hh_{\mbox{\rm\tiny ph}}$ is given by 
\begin{equation}
\label{eq-repreg}
 \bigl(\pi_{\omega}(A)\psi\bigr)_n\;=\;
  \sum_{m\in\ZM^{2}}
    \,A(\tau ^{-n} \omega, m - n)\;
\Xi^{n_1-m_1}
       \psi_m
\mbox{ , }
\hspace{1cm}
       \psi \in \Hh_{\mbox{\rm\tiny ph}}
\mbox{ . }
\end{equation}
\noindent In the following, we will again simply write $A_\omega=\pi_\omega(A)$. Then \eqref{eq-CovRel} can also be seen as a relation between different representations $\pi_{\omega}$. Furthermore, these representations are strongly continuous in $\omega$. Now $\| A\| = \sup_{\omega \in \Omega}\| A_{\omega}\|$ defines a $C^*$-norm on  $C_c(\Omega \times \ZM^{2},\mbox{\rm Mat}(2L,\CM))$ and the 
observable C$^*$-algebra $\Aa=C(\Omega) \rtimes \ZM^{2}\otimes\mbox{\rm Mat}(2L,\CM)$ is defined as the completion of $C_c(\Omega \times \ZM^{2},\mbox{\rm Mat}(2L,\CM))$  under this norm. It is also convenient to view $\Aa$ as $2\times 2$ matrices (on the particle-hole space $\CM^2_\ph$) with entries given by spinfull covariant operators. This corresponds to writing the algebra as $\Aa= C(\Omega) \rtimes \ZM^{2}\otimes\mbox{\rm Mat}(L,\CM)\otimes \mbox{\rm Mat}(2,\CM)$. 

\subsection{Derivations on $\Aa$}

On the C$^*$-algebra $\Aa$ exists a $2$-parameter group $k\in \TM^2\mapsto\rho_k$ of $*$-automorphisms defined by
$$
(\rho_k A)(\omega,n)
\;=\;
e^{-\imath k\cdot n}\,A(\omega,n)
\;,
$$
where $k\cdot n=k_1n_1+k_2n_2$, namely $\rho_k$ is linear and satisfies $\rho_k(AB)=\rho_k(A)\rho_k(B)$, $\rho_k(A^*)=\rho_k(A)^*$ and the group property $\rho_k\rho_{k'}=\rho_{k+k'}$. As every $*$-automorphism, $\rho_k$ also conserves the C$^*$-norm. Its generators ${\nabla}=(\nabla_1,\nabla_2)$ are unbounded, but closed $*$-derivations with domain $C^1 (\Aa)$. They satisfy the Leibniz rule
$$
\nabla (AB)\;=\; (\nabla A)B+A(\nabla B)
\;,
\qquad
A,B\in C^1(\Aa)
\;.
$$
Explicitly the generators are given by 
\begin{equation}
\label{eq-nablader}
 \nabla_j A (\omega, n) \;=\;
-\,  \imath\, n_j\, A (\omega, n)
\mbox{ , }
\qquad
A\in C^1 (\Aa)\;.
\end{equation}
\noindent 
Let us write out the connection with the position operator ${X}=(X_1,X_2)$ on the one-particle Hilbert space $\Hh=\ell^2(\ZM^2)\otimes\CM^L$ which as usual is defined by 
$$
(X_j \psi)_n \;= \;n_j \psi_n\;,
\qquad
\psi=(\psi_n)_{n\in\ZM^2}\in\Hh\;.
$$
Then its extension to $\Hh_\ph=\Hh\otimes \CM^2_\ph$ is by definition $X\otimes \one_2$. With this definition, one has
\begin{equation}
\label{eq-derivrep}
(\nabla_jA)_\omega
\;=\;
\imath [A_{\omega}, X_j\otimes \one_2]
\;.
\end{equation}
Let us point out that $X_j\otimes \one_2$ is {\it not} the BdG operator associated to $X_j$ which would be $X_j\otimes  Q=X_j\oplus(-X_j)$. Of course, \eqref{eq-derivrep} can also be written as $\nabla A=\imath[A,X]$.

\subsection{The trace per unit volume on $\Aa$}
\label{sec-trace}

Given a $\tau$-invariant probability measure $\PM$ on $\Omega$, a positive trace $\Tt$ on $\Aa$ (and each $\Aa$) is defined by
\begin{equation}
\label{eq-tracepervol}
 \Tt(A) \; = \;
  \int_{\Omega}
    \PM(d\omega)\;
    \TR\bigl(A(\omega , 0)\bigr)
\;=\;
\EE_\PM\,
\TR\bigl(A(\omega , 0)\bigr)
\mbox{ . }
\end{equation}
The following is readily verified.

\begin{lemma}
\label{lem-trace} $\Tt$ is a linear functional defined on all of $\Aa$ and satisfies

\vspace{.1cm}

\noindent {\rm (i)} {\rm (}normalization{\rm )} $\Tt({\bf 1}) =  2L$

\vspace{.1cm}

\noindent {\rm (ii)} {\rm (}positivity{\rm )} $\Tt(A^*A)\geq 0$ and $\Tt(A^*)=\overline{\Tt(A)}$

\vspace{.1cm}

\noindent {\rm (iii)} {\rm (}cyclicity{\rm )} $\Tt(AB)=\Tt(BA)$

\vspace{.1cm}

\noindent {\rm (iv)} {\rm (}norm bound\,{\rm )} $\Tt(|A B|)\leq \|A\|\,\Tt(|B|)$ where $|B|=(B^*B)^{\frac{1}{2}}$

\vspace{.1cm}

\noindent {\rm (v)} {\rm (}invariance{\rm )} $\Tt(\nabla A)=0$ for $A\in C^1(\Aa)$

\vspace{.1cm}

\noindent {\rm (vi)} {\rm (}partial integration{\rm )} $\Tt(A\nabla B)=-\Tt(\nabla A \,B)$ for $A,B\in C^1(\Aa)$

\end{lemma}

If $\PM$ is in addition ergodic, then Birkhoff's ergodic theorem implies that for any increasing sequence $(\Lambda_l)_{l\in\NM}$ of cubes centered at the origin
\begin{equation}
\label{eq-tracepervol2}
 \Tt(A) \; = \;
\EE_\PM\;
    \TR\bigl(\langle 0|A_{\omega}|0\rangle\bigr)
     \;   =\;
      \lim_{|\Lambda|\rightarrow\infty}
       \frac{1}{|\Lambda| }
        \sum_{n\in\Lambda\cap\ZM^2}
         \TR\bigl(\langle n| A_{\omega} |n\rangle\bigr)
\mbox{ , }
\end{equation}

\noindent for $\PM$-almost all $\omega\in\Omega$. This shows that $\Tt$ is the trace
per unit volume.

\subsection{Sub-algebra of operators with charge conservation}
\label{sec-ChargeConserv}

An operator on the BdG Hilbert space has charge conservation if and only if it commutes with the charge operator $Q$. As already pointed out in Section~\ref{sec-Charge} this simply means that the operator is diagonal. The subalgebra of $\Aa$ of such operators is denoted by $\Aa_Q$. On it acts the charge derivation defined by 
$$
\nabla^Q A
\;=\;
\imath [A,X]\,Q
\;=\;
\imath [A,XQ]
\;.
$$
Of course, there is little mathematical and physical content in all this, as BdG systems with charge conservation can simply be described by the Hamiltonian on the one-particle Hilbert space $\Hh=\ell^2(\ZM^2,\CM^L)$. In fact, this short section was merely inserted to stress the similarities with U$(1)$-invariant systems treated next.

\subsection{$\mbox{\rm U}(1)$ invariant sub-algebra and spin derivations}
\label{sec-U1Inv}

As in Section~\ref{sec-U1inv}, the $\mbox{\rm U}(1)$ invariance is taken w.r.t. rotations around the $3$-axis. Thus $\Aa_{{\rm U}(1)}$ is defined as the subset of operators in $\Aa$ for which the commutation relation $j=3$ in \eqref{eq-SU2inv} holds. As $s^3$ and thus $S^3$ defined in \eqref{eq-spinsymbol2} is symmetric, this definition can be written as 
$$
\Aa_{{\rm U}(1)}
\;=\;
\bigl\{\,A\in\Aa\,:\,[A,S^3]=0\,\bigr\}
\;.
$$
This implies that $\Aa_{{\rm U}(1)}$ is a $*$-subalgebra of $\Aa$. The same proof as in Propositions~\ref{prop-SU2inv} and \ref{prop-U1inv} leads to the following result.

\begin{proposi}
\label{prop-algU1inv}
The algebra $\Aa_{{\rm U}(1)}\subset\Aa$ has the representation
$$
\Aa_{{\rm U}(1)}
\;=\;
\left\{\,
\begin{pmatrix}
a & b \\ c & d
\end{pmatrix}
\,:\,a,d\;\mbox{\rm diagonal and }b,c\;\mbox{\rm cross-diagonal in spin degree of freedom }\,\right\}
\;.
$$
Here the $2\times 2$ matrix is in the particle-hole space and the entries are covariant operators $a,b,c,d\in C(\Omega) \rtimes \ZM^2\otimes\mbox{\rm Mat}(L,\CM)$. 
\end{proposi}



Next let us set $C^1(\Aa_{{\rm U}(1)})=C^1(\Aa)\cap \Aa_{{\rm U}(1)}$. For $A\in C^1(\Aa_{{\rm U}(1)})$ one can now define the spin derivation (along the $3$-direction of the spin)
$$
\nabla^{S^3}A
\;=\;
\nabla A\;S^3\;=\;S^3\; \nabla A
\;.
$$
Note that this is indeed a $*$-derivation on $\Aa_{{\rm U}(1)}$.
In the representation one has
\begin{equation}
\label{eq-spinder}
(\nabla^{S^3}A)_\omega
\;=\;
\imath [A_{\omega},X\,S^3]
\;=\;
\imath [A_{\omega}, X\otimes \one_L\otimes \one_2]\;S^3
\;.
\end{equation}
This follows from the decomposition $X\,S^3 =(X\otimes \one_2)\; S^3$ and the fact that $A$ commutes with $S^3$. 

\subsection{$\mbox{\rm SU}(2)$ invariant sub-algebra}

Also the definition of $\mbox{\rm SU}(2)$-invariant operators is motivated by the corresponding invariance of BdG operators given in Section~\ref{sec-SU2inv}:
$$
\Aa_{{\rm SU}(2)}
\;=\;
\bigl\{\,A\in\Aa\,:\,[A,S^1]=[A,S^2]=0\;,\;\;[A,S^3]=0\,\bigr\}
\;.
$$
This is a $*$-subalgebra of $\Aa$.  Following the argument of Proposition~\ref{prop-SU2inv} leads to

\begin{proposi}
\label{prop-algSU2inv}
The algebra $\Aa_{{\rm SU}(2)}\subset\Aa$ has the representation
$$
\Aa_{{\rm SU}(2)}
\;=\;
\left\{\,
\begin{pmatrix}
a\otimes\one_L & b\otimes T \\ c\otimes T & d\otimes\one_L
\end{pmatrix}
\,:\,a,b,c,d\in C(\Omega) \rtimes \ZM^2
\,\right\}
\;,
$$
where $T$ is the $L\times L$ cross-diagonal matrix introduce in {\rm Section~\ref{sec-SU2inv}}. 
\end{proposi}

The algebra $\Aa_{{\rm SU}(2)}$ is invariant under functional calculus, namely if $A=A^*\in \Aa_{{\rm SU}(2)}$ and $f:\RM\to\CM$, then $f(A)\in \Aa_{{\rm SU}(2)}$.  Let us point out that $\nabla^{S^3}A$ is well-defined for $A\in \Aa_{{\rm SU}(2)}\subset \Aa_{{\rm U}(1)}$, but it is not true that $\nabla^{S^3}A\in\Aa_{{\rm SU}(2)}$.

\section{Topological invariants of BdG Hamiltonians}
\label{sec-invariants}

In this section, we explain how to calculate the invariants listed in the table in Section~\ref{sec-classes}. These numbers are all topological invariants that do not change as one changes the Hamiltonian or the chemical potential, provided that the mobility gap does not close and the stated symmetries are conserved. Hence they can be considered as labels for different phases of the systems. The physical phenomena associated to these invariants will be discussed in the remainder of the paper.

\vspace{.2cm}

The main invariant is the (two-dimensional, sometimes also called first) Chern number of the BdG Fermi projection $P=\chi(H\leq 0)$ defined by (see \cite{BES,PS}, also \cite{ASS})
$$
\Ch(P)
\;=\;
2\pi\imath\,\Tt(P[\nabla_1 P,\nabla_2 P])
\;.
$$
It is well-defined as long as there is dynamical localization at the Fermi level, namely the bound
\begin{equation}
\label{eq-loccond}
\Tt\bigl(
|\nabla  P|^2 \bigr)
\;<\;\infty
\end{equation}
holds. One particular case of this is, of course, that $H$ has a spectral gap at $0$. A proof of \eqref{eq-loccond} for random BdG Hamiltonians with closed central is an interesting problem which, to our knowledge, has not been dealt with. The techniques of \cite{GM,DDS} only cover band edges of $H$, but not $E=0$ for a closed bulk gap. Under the localization condition \eqref{eq-loccond}, there is an index theorem \cite{BES,PS} linking the Chern number to the Noether index of a Fredholm operator
$$
\Ch(P)\;=\;\Ind(P_\omega FP_\omega)
\;,
$$
where $F=(X_1+\imath X_2)|X_1+\imath X_2|^{-1}$ is the so-called Dirac phase. The r.h.s. is actually a random integer number, but for a homogeneous Hamiltonian it is known to be $\PM$-almost surely constant. Thus $\Ch(P)\in\ZM$, giving the entry in the table for Class D. As to the Class DIII, the $\ZM_2$ index appearing in the table in Section~\ref{sec-classes} can be obtained as the parity of the kernel dimension of the Fredholm operator $P_\omega FP_\omega$, see \cite{GS}. In this paper, we will not further discuss the physics associated to a non-trivial value of this $\ZM_2$ invariant, but rather focus on the integer valued invariants which will be directly linked to response coefficients. Let us now analyze the properties of $\Ch(P)$ whenever the Hamiltonian has supplementary symmetries corresponding to other cases in the table.

\begin{proposi}
\label{prop-TopoChargeCon}
Let $H=\diag(h-\mu,-\overline{h}+\mu)$ be a {\rm BdG} Hamiltonian having charge conservation. Then 
$$
{\rm Ch}(P)\;=\;2\,\Ch(p)
\;,
$$
where $p=\chi(h\leq \mu)$ is the Fermi projection of the particle sector only.
\end{proposi}

\noindent {\bf Proof:} As $P=p\oplus (\one-\overline{p})$ and the Chern number is additive     the identity
%
$$
\Ch(\one-\overline{p})
\;=\;
-\;\Ch(\overline{p})
\;=\;
\Ch(p)
\;,
$$
implies the result.
\hfill $\Box$

\vspace{.2cm}

The following proposition is obtained in a similar manner.

\begin{proposi}
\label{prop-topU(1)}
Let $H$ be {\rm BdG} Hamiltonian with spin $s=\frac{L-1}{2}$ and $\mbox{\rm U}(1)$ invariance.  Then
$$
{\rm Ch}(P)\;=\;\sum_{l=1}^L {\rm Ch}(P_l)\;,
$$
where $P_l=\chi(H_\redu^{(l)}\leq 0)$ is the Fermi projection of the reduced Hamiltonian  $H_\redu^{(l)}$ appearing in the decomposition $H=\bigoplus_{l=1}^LH_\redu^{(l)}$ given in {\rm Proposition \ref{prop-U1inv}}.
\end{proposi}

\begin{proposi}
\label{prop-topSU(2)}
Let $H$ be a {\rm BdG} Hamiltonian with half-integer spin $s=\frac{L-1}{2}$ and $\mbox{\rm SU}(2)$ invariance.  Then
$$
{\rm Ch}(P)\;=\;L\; {\rm Ch}(P_\redu)\;,
$$
where $P_\redu=\chi(H_\redu)$ is the Fermi projection of the reduced Hamiltonian $H_\redu$ as given in {\rm Proposition~\ref{prop-SU2inv}}, with $\sigma=1$. Moreover, ${\rm Ch}(P_\redu)\in 2\,\ZM$ so that $\Ch(P)\in 2L\,\ZM$.
\end{proposi}

\noindent {\bf Proof:} By Proposition~\ref{prop-SU2inv}, $P=\oplus_{l=1,\ldots,\frac{L}{2}}P_\redu\oplus P'_\redu$ where $P'_\redu=\chi(H'_\redu)$ is the Fermi projection of the second reduced Hamiltonian. As both of them are unitarily equivalent by a local unitary, $\Ch(P'_\redu)=\Ch(P_\redu)$. This implies the first claim. The second one follows from the odd PHS of $H_\redu$, see \cite{DS} or \cite{GS}.
\hfill $\Box$

\begin{proposi}
\label{prop-ChTRS}
Let $H$ be an {\rm BdG} Hamiltonian which also has {\rm TRS}. Then $\Ch(P)=0$.
\end{proposi}

\noindent {\bf Proof.} The TRS of $H$ implies that $(R\oplus R)\,P\,(R\oplus R)^*=\overline{P}$, see Section~\ref{sec-TRS}. As $\Tt$ is unitarily invariant under $R\oplus R$, one deduces $\Ch(P)=-\overline{\Ch(P)}=-\Ch(P)$.
\hfill $\Box$

\vspace{.2cm}

To conclude this section, let us cite ({\it e.g.} from \cite{DDS}) some examples with non-trivial Chern numbers by referring to the list in Section~\ref{sec-interaction}. For the $p\pm\imath p$ wave superconductor, the Chern number $\Ch(P)$ is equal to $\pm 1$ or $0$, pending on the values of $\mu$ and $\delta$. This is considered to be the most elementary Class D model with a non-trivial invariant and actually, written in Fourier space, the projection $P$ is essentially the Bott projection. A non-trivial example for Class C is the $d\pm\imath d$ wave superconductor. The Chern number of the reduced Fermi projection $P_\redu$ of Proposition~\ref{prop-topSU(2)} is equal to $\pm 2$ or $0$, again with values depending on $\mu$ and $\delta$.

\section{Hall conductance and BBC for BdG Hamiltonians}
\label{sec-BBC}

This section transposes the main facts on the BBC for two-dimensional systems of independent fermions, in the form proved in \cite{KRS,PS}, to the framework of BdG Hamiltonians $H$ on the Hilbert space $\ell^2(\ZM^2,\CM^{2L})$. In the bulk, this will involve the Hall matter currents resulting from a gravitational field and on the boundary simply the persistent matter currents. Both effects are connected by the BBC. While these are possibly merely theoretical developments with no experimental implications, they serve as mathematical preparation for the following sections. We will assume to be in the framework of homogeneous operators described in Section~\ref{sec-alg}, hence $H\in\Aa$.

\vspace{.2cm}

Let us begin by deriving the Kubo formula for the Hall matter currents by applying the general strategy outlined in Appendix~\ref{app-kubo}. The matter current operator is, according to Appendix~\ref{app-MatterCur}, given by 
$$
\Jj
\;=\;
\imath [H,X]
\;=\;
\nabla H
\;.
$$
By a standard Fourier argument \cite{BES,PS}, there is no current when the system is in a Fermi-Dirac state:
\begin{equation}
\label{eq-TraceInva}
\Tt\bigl(f_{\beta}(H)\,\nabla H \bigr)
\;=\;0\;.
\end{equation}
Now let us add a perturbation by a linearly growing gravitational field. Hence, in the notation of Appendix~\ref{app-kubo}, the perturbation is $\lambda\Pp$ with coupling constant $\lambda$ and $ \Pp=X_2$ acting on $\Hh_\ph$. The current operator in Appendix~\ref{app-kubo} is $\nabla_1H$. By \eqref{eq-TraceInva}, the current at equilibrium vanishes. In conclusion, all the hypothesis of Appendix~\ref{app-kubo} are satisfied and the Kubo formula for the matter Hall conductance is given by
\begin{equation}
\label{eq-MatterHallKubo}
\sigma(\beta,\delta)
\; = \;
\frac{1}{2}\;\Tt\bigl(
\nabla_2f_{\beta}(H)
\;
(\delta+\Ll_H)^{-1}(\nabla_1H)
 \bigr)
\;.
\end{equation}
The focus is now on the zero temperature case $\beta=\infty$ for which $f_{\infty}(H)=\chi(H\leq 0)=P$ is the BdG Fermi projection (the dependence of $H$ and $P$ on $\mu$ as given in \eqref{eq-BdGfirst}  is suppressed). The exchange of the limits $\delta\to 0$ and $\beta\to\infty$ on the Kubo formula \eqref{eq-MatterHallKubo} can be analyzed as in \cite{BES} and \cite[Theorem~3]{ST}, provided that the density of states has no atom at the Fermi level $E=0$ and that the following localization condition \eqref{eq-loccond} holds for the BdG Fermi projection $P$. 

\begin{theo}
\label{theo-MatterHallKuboChern}
Let $H$ be a {\rm BdG} Hamiltonian. Suppose that the localization condition {\rm \eqref{eq-loccond}} holds and that $H$ has no infinitely degenerate eigenvalue at $E=0$. Then the Kubo-Chern formula for the matter Hall conductance at zero temperature holds:
\begin{equation}
\label{eq-MatterHallKuboChern}
\sigma
\; = \;
\frac{1}{2}\;\imath\;
\Tt\bigl(
P\;[\nabla_1P,\nabla_2P]
 \bigr)
\;=\;
\frac{1}{2}\;\frac{1}{2\pi}\;\Ch(P)
\;.
\end{equation}
Suppose that $I\subset\RM$ is an interval such that the localization condition {\rm \eqref{eq-loccond}} holds for all $P=P_\mu$ with $\mu\in I$ and that the {\rm DOS} has no atoms in $I$. Then the matter Hall conductance $\mu\in I\mapsto\sigma(\mu)$ is constant and takes values in the set $\frac{1}{4\pi}\,\ZM$.
\end{theo}

\noindent {\bf Proof.} 
At zero temperature, the matter Hall conductance \eqref{eq-MatterHallKubo} is given by
$$
\sigma
\; = \;
\lim_{\delta\downarrow 0}\;
\frac{1}{2}\;\Tt\bigl(
\nabla_2 P
\;
(\delta+\Ll_H)^{-1}(\nabla_1H)
 \bigr)
 \;.
 $$
Now
$$
\nabla_2 P
\;=\;
P\, \nabla_2 P\;(\one-P)
\,+\,
(\one-P)\,\nabla_2 P \,P
\;,
$$
and, due to the detailed proof in \cite[Theorem~3]{ST},
\begin{eqnarray*}
\lim_{\delta\downarrow 0}\;
P \,(\delta+\Ll_H)^{-1}(\nabla_1H)\,(\one-P)
& = &
\imath\,P\;\nabla_1P\,(\one-P)
\;,
\\
\lim_{\delta\downarrow 0}\;
(\one-P)\,(\delta+\Ll_H)^{-1}(\nabla_1H)\,P
& = &
-\,\imath\,(\one-P)\,\nabla_1P\,P
\;.
\end{eqnarray*}
This implies the first claim. The second one holds because the techniques of \cite{BES} transpose directly to the present situation. 
\hfill $\Box$

\vspace{.2cm}

Next let us introduce and study the matter currents in a half-space. The half-space Hamiltonian $\widehat{H}$ is given, for simplicity, just by the restriction of $H$ to the halfspace $\ell^2(\ZM\times\NM)\otimes\CM^{2L}$ (Dirichlet boundary conditions, but actually all the below is valid for any local boundary condition \cite{KRS,PS}).  The observable for matter edge currents along the boundary is the $1$-component $\widehat{\Jj}_1=\imath[\widehat{H},X_1]$ of the half-space matter current operator. Let $g(\widehat{H})$ be a density matrix defined by spectral calculus from a real-valued, smooth and even function $g$ on $\RM$ with $\int dE\,g(E)=1$. We suppose that supp$(g)\cap\sigma(H)=\emptyset$, notably that $g$ is supported by a central (bulk) gap of $H$. In this situation the edge current $\hat{\jmath}(g)$ of state $g(\widehat{H})$ is defined by
\begin{equation}
\label{eq-edgecurrentdef}
\hat{\jmath}(g)
\;=\;
-\;\frac{1}{2}\;
\widehat{\Tt}\bigl(g(\widehat{H})\,\widehat{\Jj}_1\bigr)
\;=\;
-\;\frac{1}{2}\;
\widehat{\Tt}\bigl(g(\widehat{H})\,\nabla_1\widehat{H}\bigr)
\;,
\end{equation}
where $\widehat{\Tt}=\Tt_1\,\Tr_2$ is the trace per unit volume in the $1$-direction along the boundary and $\Tr_2$ the usual trace in the $2$-direction perpendicular to the boundary. For any operator family $\widehat{A}=(\widehat{A}_\omega)_{\omega\in\Omega}$ on $\ell^2(\ZM\times\NM)\otimes\CM^{2L}$ which is homogeneous in the $1$-direction, it is more formally defined by
\begin{equation}
\label{eq-Thatdef}
\widehat{\Tt}(\widehat{A})
\;=\;
\EE_\PM \sum_{n_2\geq 0}\;\Tr\,\langle 0,n_2|\widehat{A}_\omega|0,n_2\rangle
\;.
\end{equation}
This defines a {\it bona fide} trace on the Toeplitz extension of $\Aa$ which contains the half-space restrictions of homogeneous operators \cite{KRS,PS}. As is obvious from this definition, not all operators $\widehat{A}$ are traceclass w.r.t. $\widehat{\Tt}$ due to the sum over $n_2$. Thus one has to prove that \eqref{eq-edgecurrentdef} actually makes sense for adequate functions $g$. This is part of the result below. Let us also add that the factor $\frac{1}{2}$ in \eqref{eq-edgecurrentdef} stems from \eqref{eq-Fac1/2}. Furthermore, the sign can heuristically be explained as follows. The confinement to a half-space can be seen as gravitational potential which is infinite for negative $x_2$. This leads to a potential drop instead of a potential raise in the linear response theory above. To obtain a relation between the responses (induced currents) not containing a sign, one should add the sign in \eqref{eq-edgecurrentdef}.

\begin{theo}[\rm \cite{KRS,PS}, see also \cite{EG,BCR,MT}]
\label{theo-edgequant} 
Let $\Delta$ be a {\rm (}bulk{\rm )} gap of the {\rm (}almost sure{\rm )} spectrum of $H\in C^1(\Aa)$. Then for any positive smooth function $g$ with $\mbox{\rm supp}(g)\subset \Delta$ and $\int dE\,g(E)=1$, the operator $g(\widehat{H})$ is $\widehat{\Tt}$-traceclass and one has
\begin{equation}
\label{eq-boundaryquant}
\hat{\jmath}(g)\;=\;
\frac{1}{2}\;\frac{1}{2\pi}\;\imath\;\widehat{\Tt}
\left((U(g)^*-\one)\nabla_1 U(g)\right)
\;,
\end{equation}
namely it is $\frac{1}{4\pi}$ times the non-commutative winding number of
$$
\widehat{U}_\omega(g)\;=\;\exp\bigl(-\,2\,\pi\,\imath\,G(\widehat{H}_\omega)\bigr)
\;,
$$
where $G(E)=\int^E_{-\infty}dE'\,g(E')$. This winding number is also equal to the $\PM$-almost sure index of the Fredholm operator ${\Pi}_1^*\widehat{U}_\omega(g)^*{\Pi}_1$ where ${\Pi}_1$ is the embedding of the quarter plane Hilbert space $\ell^2(\NM\times\NM)\otimes\CM^{2L}$ into  $\ell^2(\ZM\times\NM)\otimes\CM^{2L}$. This integer is independent of the choice of boundary conditions and of the choice of the function $g$. The {\rm BBC} now states that this integer is also linked to the bulk invariant: 
\begin{equation}
\label{eq-BoundCur}
\hat{\jmath}(g)\;=\;
\frac{1}{2}\;\frac{1}{2\pi}\;\Ch(P)
\;.
\end{equation}
\end{theo}

One of the claims of the theorem states that $\hat{\jmath}(g)$ is independent of the detailed choice of the even smooth function $g$. Hence we also call $\hat{\jmath}(g)$ the edge Hall conductance and simply denote it by $\hat{\sigma}$. Then Theorem~\ref{theo-edgequant} implies $\hat{\sigma}=\sigma$. The reader may on first sight be irritated by the factors $\frac{1}{2}$ in \eqref{eq-boundaryquant} and \eqref{eq-BoundCur}, but this will be further elucidated below. For a final comment,  let $g$ converge to $\frac{1}{|\Delta|}\chi_\Delta$ where $\Delta=[-\delta,\delta]$ is an interval lying in the central gap of $H$. Then the formula \eqref{eq-BoundCur} becomes 
\begin{equation}
\label{eq-BoundCur2}
-\;\frac{1}{2}\;\widehat{\Tt}\bigl(\chi_\Delta(\widehat{H})\,\nabla_1\widehat{H}\bigr)
\;=\;
|\Delta|\;\frac{1}{2}\;\frac{1}{2\pi}\;
\Ch(P)
\;=\;
|\Delta|\;\sigma
\;.
\end{equation}
This directly leads to a physical interpretation described in the next section.

\section{Bulk and boundary under charge conservation}
\label{sec-ChargeCons}

According to Section~\ref{sec-Charge}, a BdG Hamiltonian has charge conservation if and only if $H=\diag(h-\mu,-\overline{h}+\mu)$. Then $P=\diag(p,\one-\overline{p})$ where $p=\chi(h\leq\mu)$ and by Proposition~\ref{prop-TopoChargeCon} one has $\Ch(P)=2\,\Ch(p)$. The physical effect associated with this invariant is the (electrical) Hall conductance and its topological quantization is precisely the quantum Hall effect, {\it e.g.} \cite{BES} for a mathematical description in the spirit of the present work. Let us briefly describe within the BdG formalism how to obtain the connection between Hall conductance and Chern number. The electron system is submitted to a constant electric field leading to a perturbation $\Pp=\lambda X_2\,Q$  in the linear response theory as described in Appendix~\ref{app-kubo} where $\lambda=q\,\Ee$ contains the particle charge $q$ and the electric field $\Ee$. One then measures the charge current $\Jj^Q_1=\nabla_1 H\,Q=\nabla_1^Q H$ in the $1$-direction. By the same argument as in Section~\ref{sec-BBC} (and in \cite{BES}) the Kubo formula leads to a zero temperature response coefficient
$$
\sigma_Q
\;=\;
\frac{1}{2}\;\imath\;
\Tt\bigl(
P\;[\nabla^Q_1P,\nabla^Q_2P]
 \bigr)
\;.
$$
Now $\Tt$ is invariant under $Q$ and by charge conservation $[P,Q]=0$. As $Q^2=\one$, the two factors $Q$ therefore cancel out and one obtains the well-known formula with a factor $\frac{1}{2}$:
$$
\sigma_Q
\;=\;
\frac{1}{2}\;\frac{1}{2\pi}\;\Ch(P)
\;=\;
\frac{1}{2\pi}\;\Ch(p)
\;.
$$
This result is consistent with the fact under the charge conservation the BdG formalism is just a rewriting of the theory for the single electron Hamiltonian $h-\mu$.

\vspace{.2cm}

Next let us study the electric boundary currents. Comparing with \eqref{eq-edgecurrentdef}, one may be tempted to consider $-\frac{1}{2}\,\widehat{\Tt}\bigl(g(\widehat{H})\,\widehat{\Jj}_{Q,1}\bigr)$, but this quantity actually vanishes. Actually, as the system has charge conservation, the charge currents coincide with the matter boundary currents as defined in \eqref{eq-edgecurrentdef}, namely one has to consider $-\frac{1}{2}\,\widehat{\Tt}\bigl(g(\widehat{H})\,Q\,\widehat{\Jj}_{Q,1}\bigr)$ and the two factors $Q$ cancel out as in the bulk theory. They are calculated from the half-space Hamiltonian $\widehat{H}=\diag(\hat{h}-\mu,-\overline{\hat{h}}+\mu)$ and thus satisfy
$$
\hat{\jmath}_Q(g)
\;=\;
-\;\frac{1}{2}\;
\left(
\widehat{\Tt}\bigl(g(\hat{h}-\mu)\,\nabla_1 \hat{h}\bigr)
\;+\;
\widehat{\Tt}\bigl(g(-\overline{\hat{h}}+\mu)\,\nabla_1(-\overline{\hat{h}})\bigr)
\right)
\;=\;
-\,\widehat{\Tt}\bigl(g(\hat{h}-\mu)\,\nabla_1 \hat{h}\bigr)
\;.
$$
Hence  \eqref{eq-BoundCur} simply becomes 
$$
-\;\widehat{\Tt}\bigl(g(\hat{h}-\mu)\,\nabla_1 \hat{h}\bigr)
\;=\;
\frac{1}{2\pi}\;\Ch(p)
\;,
$$
and \eqref{eq-BoundCur2}
\begin{equation}
\label{eq-BBCx}
-\;\widehat{\Tt}\bigl(\chi_\Delta(\hat{h}-\mu)\,\nabla_1 \hat{h}\bigr)
\;=\;
|\Delta|\;\frac{1}{2\pi}\;\Ch(p)
\;.
\end{equation}
As a preparation for the thermal edge currents discussed in Section~\ref{sec-THE}, let us give some further physical intuition on \eqref{eq-BBCx} and thus the main result of Theorem~\ref{theo-edgequant}, by following \cite{SKR}. This will also explain why it is reasonable to call the mathematical object $\hat{\jmath}_Q(g)$ the edge  Hall conductance $\hat{\sigma}_Q$, see \cite{SKR,KRS}. For this purpose, let us write $\chi_\Delta(\hat{h}-\mu)=\hat{p}_+-\hat{p}_-$ where $\hat{p}_\pm=\chi(\hat{h}\leq \mu_\pm)$ are the Fermi projections associated to $\mu_\pm=\mu\pm\delta$ where $\Delta=[-\delta,\delta]$. One obtains formally
\begin{equation}
\label{eq-BBC}
\widehat{\Tt}\bigl(
\hat{p}_-\, \nabla_1\hat{h}\bigr)
\;-\;
\widehat{\Tt}\bigl(\hat{p}_+\, \nabla_1\hat{h}\bigr)
\;=\;
(\mu_+-\mu_-)\;\frac{1}{2\pi}\;\Ch(p)
\;.
\end{equation}
This is mathematically not sound because neither $ \hat{p}_+$ nor $\hat{p}_-$ is traceclass w.r.t. $\widehat{\Tt}$, but only the difference is (for $\Delta$ lying in a gap of $H$). On the other hand, now the following interpretation becomes apparent. Suppose we consider a Hall bar which has two different chemical potentials $\mu_+$ and $\mu_-$ at the upper and lower edge of the bar. Then the first term on the l.h.s. of \eqref{eq-BBC} is the chiral edge current flowing along the lower edge of a sample, and the second contribution stems from the opposite upper edge where the chiral current flows in the opposite direction, thus leading to a different sign. Calculating these two contributions separately assumes that there is no tunnel effect from upper to lower edge. Resuming, the l.h.s. of \eqref{eq-BBC} is the total current flowing along the two edges of the system. This current is, according to  \eqref{eq-BBC}, equal to the potential difference $\mu_+-\mu_-$ times the bulk Hall conductivity $\sigma=\frac{1}{2\pi}\,\Ch(p)$.

\section{Spin quantum Hall effect}
\label{sec-SQHKubo}

In the spin Hall effect, one measures one component of the spin current in a direction transverse to the gradient of a constantly growing magnetic Zeeman field. For this to make sense, this component of the spin needs to be conserved. We choose it to be the $3$-component and assume throughout this section that the BdG Hamiltonian $H$ satisfies the conservation property $[H,S^3]=0$, which is also called a U$(1)$ invariance, see Section~\ref{sec-U1Inv}. This conservation law replaces the charge conservation in the conventional (electrical) Hall effect discussed in Section~\ref{sec-ChargeCons}. Thus there are a several similarities, and in a sense made more precise below the spin quantum Hall effect is merely a direct sum of several quantum Hall effects.  In particular, we will derive a Kubo formula for the spin Hall effect and show that it is a topological quantity at zero temperature and in presence of a mobility gap, namely it is connected to Chern numbers. For the special case of an $\mbox{\rm SU}(2)$ invariance, one, moreover, knows that the Chern number is even. Furthermore, we also provide a purely quantum mechanical calculation of the spin edge currents in presence of a bulk gap. There are numerous physics papers on the spin quantum Hall effect. One of the first ones is \cite{SMF} which considered the $\mbox{\rm SU}(2)$ invariant singlet $d\pm \imath d$ superconductors. A corresponding network model is studied in \cite{GLR}.

\vspace{.2cm}

Let us begin by deriving the Kubo formula by applying once again the general strategy outlined in Appendix~\ref{app-kubo}. The spin current operator for the $3$-component is, according to Appendix~\ref{app-SpinCurr}, given by 
$$
\Jj_{S^3}
\;=\;
\imath [X,H]\,S^3
\;=\;
\imath [XS^3,H]
\;=\;
\nabla^{S^3} H
\;.
$$
At equilibrium there is  no spin current, namely $\Tt\bigl(f_{\beta}(H)\,\nabla^{S^3}H \bigr)=0$. Now let us add a perturbation by a linearly growing magnetic field $B_3$ (say growing in the spacial direction $2$) which is only coupled to the spin degree of freedom (Zeeman field). Hence, in the notation of Appendix~\ref{app-kubo}, the perturbation is $\lambda\Pp$ with coupling constant $\lambda=B_3$ and $ \Pp=X_2 S^3$ acting on $\Hh_\ph$. Note that $\Ll_\Pp=\nabla^{S^3}_2$ is indeed the generator of an automorphism group on $\Aa_{{\rm U}(1)}$, and thus by a Dyson series argument the time-evolution $\Ll_H+\lambda\Ll_\Pp$ is well-defined on $\Aa_{{\rm U}(1)}$. One is interested in calculating the $3$-component of the  spin current in the $1$-direction, namely $\nabla^{S^3}_1H$. As the spin current at equilibrium vanishes, all the hypothesis of Appendix~\ref{app-kubo} are satisfied and the Kubo formula for the spin Hall conductance is given by
\begin{equation}
\label{eq-spinHallKubo}
\sigma_{S^3}(\beta,\delta)
\; = \;
\frac{1}{2}\;\Tt\bigl(
\nabla^{S^3}_2f_{\beta}(H)
\;
(\delta+\Ll_H)^{-1}(\nabla^{S^3}_1H)
 \bigr)
\;.
\end{equation}
Now  the $\beta\to\infty$ and $\delta\to0 $ limits can be dealt with just as in Section~\ref{sec-BBC}. 

\begin{theo}
\label{theo-spinHallKuboChern}
Let $H$ be an $\mbox{\rm U}(1)$ invariant {\rm BdG} Hamiltonian. Suppose that the localization condition {\rm \eqref{eq-loccond}} holds and that $H$ has no infinitely degenerate eigenvalue at $E=0$. Then the Kubo-Chern formula for the spin Hall conductance at zero temperature holds:
\begin{equation}
\label{eq-spinHallKuboChern}
\sigma_{S^3}
\; = \;
\frac{1}{2}\;\imath\;
\Tt\bigl(
P\;[\nabla^{S^3}_1P,\nabla^{S^3}_2P]
 \bigr)
\;.
\end{equation}
Furthermore, in terms of the Chern numbers $\Ch(P_l)$ described in {\rm Proposition~\ref{prop-topU(1)}},
$$
\sigma_{S^3}
\;=\;
\frac{1}{16\pi}\;
\sum_{l=1}^L (L+1-2l)^2\;{\rm Ch}(P_l)\;.
$$
In particular, $\sigma_{S^3}\in\frac{1}{16\pi}\,\ZM$. If $H$ has $\mbox{\rm SU}(2)$ invariance, then
$$
\sigma_{S^3}
\;=\;
\frac{L(L^2-1)}{48\pi}\;{\rm Ch}(P_\redu)\;,
$$
with $\Ch(P_\redu)\in 2\,\ZM$. Thus for spin $s=\frac{1}{2}$, one has $\sigma_{S^3}\in\frac{1}{4\pi}\,\ZM$.
\end{theo}

\noindent {\bf Proof.} The derivation of \eqref{eq-spinHallKuboChern} follows the arguments of Theorem~\ref{theo-MatterHallKuboChern}. For the second identity, one proceeds as in Proposition~\ref{prop-topU(1)}, based on  Proposition~\ref{prop-U1inv}, by invoking Appendix~\ref{app-su2} for the calculation of $(S^3)^2$. Finally, when $H$ has  $\mbox{\rm SU}(2)$ invariance, then $\Ch(P_l)=\Ch(P_\redu)\in 2\,\ZM$ by the proof of Proposition~\ref{prop-topSU(2)}. Carrying out the sum over $l$ implies the result.
\hfill $\Box$

\vspace{.2cm}

Also the local constancy of $\sigma_{S^3}$ under variation of the chemical potential can be stated and proved exactly as in Theorem~\ref{theo-MatterHallKuboChern}, but this is not spelled out here. Similarly, as for the charge boundary currents, the spin boundary currents are now defined by
$$
\hat{\jmath}_{S^3}(g)
\;=\;
-\;\frac{1}{2}\;
\widehat{\Tt}\bigl(g(\widehat{H}){S^3}\,\widehat{\Jj}_{S^3}\bigr)
\;.
$$
Due to the U$(1)$ invariance, this becomes
$$
\hat{\jmath}_{S^3}(g)
\;=\;
-\;\frac{1}{2}\;
\widehat{\Tt}\bigl(g(\widehat{H})(S^3)^2\,\widehat{\Jj}\bigr)
\;.
$$
Now one can go into the eigenbasis of $(S^3)^2$ and use the BBC in each of its eigenspaces. Then combining this with the result of Theorem~\ref{theo-spinHallKuboChern}, one obtains 
$$
\hat{\jmath}_{S^3}(g)
\;=\;
\sigma_{S^3}
\;,
$$
which can again be interpreted as the equality of edge spin conductance $\hat{\sigma}_{S^3}$ with the bulk spin conductance $\sigma_{S^3}$.


\section{Thermal quantum Hall effect}
\label{sec-THE}

When a system is submitted to a temperature gradient, often modeled by a gravitational field \cite{Lut}, it may result in a heat current in a transverse direction. This  phenomenon is called the thermal Hall effect, sometimes also Leduc-Righi effect, and it can be dealt with a refined version of the linear response theory \cite{Lut,SSt,OS,CHR,QSN}. Quantization of (temperature coefficient of) the thermal Hall conductance for two-dimensional BdG Hamiltonians has been shown in several works \cite{SFi,VMFT,Vis,SF}.  The derivation of the Kubo formula for the thermal Hall conductance $\kappa(\beta)$ given in \cite{SSt,VMFT,QSN,SF} is, in our opinion, not mathematical sound in a tight-binding framework and is subject to further examinations. On the other hand, we believe in the validity of the result and  thus use it as the starting point for the connection with topological invariants. In the formalism of the present paper,  the Kubo formula of \cite{SSt,VMFT,SF} reads
$$
\kappa(\beta)
\;=\;
-\;
\frac{1}{2}\;\beta\;\int_\RM dE\,E^2\,f'_\beta(E)
\;\imath\,\Tt
\big(P_E[\nabla_1P_E,\nabla_2 P_E]\big)
\;,
$$
where $P_E=\chi(H\leq E)$. At low temperature, the main contribution to the integral comes from small values of $E$ for which $2\pi\imath\,\Tt\big(P_E[\nabla_1P_E,\nabla_2 P_E]\big)=\Ch(P_E)$ is constant and equal to $\Ch(P)$ by Theorem~\ref{theo-MatterHallKuboChern}. Using the (weak form of the) Sommerfeld expansion
\begin{equation}
\label{eq-Sommerfeld}
f'_\beta(E)
\;=\;
-\;\delta(E)
\;-\;\frac{\pi^2}{6}\;T^2\;
\delta''(E)
\;+\;\Oo(T^4)
\;,
\end{equation}
one therefore deduces
\begin{equation}
\label{eq-ThermalEdgeQuant0}
\kappa(\beta)
\;\approx\;
\frac{\pi}{12}\;T\;\Ch(P)
\;.
\end{equation}
Let us stress that this holds for all two-dimensional BdG Hamiltonians with a mobility gap, in particular, also for systems with conserved charge (CAZ Class A). For these latter systems, Proposition~\ref{prop-TopoChargeCon} then shows ${\rm Ch}(P)=2\,\Ch(p)$ so that $\kappa(\beta)\approx\frac{\pi}{6}\,T\,\Ch(p)$. Furthermore, there is then also an associated Wiedemann-Franz law for the quotient of thermal and electric Hall conductivities: 
$$
\frac{\kappa(\beta)}{\sigma_Q}
\;=\;
\frac{\pi^2}{3}\,T
\;.
$$

\vspace{.2cm}

Let us now analyze the associated thermal boundary currents.  A brief discussion of these currents can be found in \cite{Kit0}, and an effective field theoretic description in \cite{SYN,NRN}. To deduce the formula for the thermal edge current density, let us adapt \eqref{eq-BBC}. According to Appendix~\ref{sec-EnergyCur}, the heat current operator is $\frac{1}{2}\,\imath[\widehat{H}^2,X_1]=\frac{1}{2}\,\nabla_1 \widehat{H}^2$. Instead of two different local chemical potentials $\mu_+$ and $\mu_-$, one now rather has two different local temperatures $\beta_+$ and $\beta_-$. For sake of simplicity, we will set $\beta_-=\infty$ and thus we have $\beta_+=\beta=\frac{1}{T}$ as the only parameter. The net thermal edge current density is then
\begin{align}
\label{eq-BBC2}
\hat{\jmath}_H(\beta)
&
\;=\;
-\;\frac{1}{2}\;
\widehat{\Tt}\bigl(f_{\beta}(\widehat{H})\,\frac{1}{2}\,\nabla_1\widehat{H}^2\bigr)
\;+\;
\frac{1}{2}\;\widehat{\Tt}\bigl(f_{\infty}(\widehat{H})\,\frac{1}{2}\,\nabla_1\widehat{H}^2\bigr)
\\
&
\;=\;
-\,\frac{1}{2}\;\widehat{\Tt}\bigl( ( f_{\beta}(\widehat{H})-f_{\infty}(\widehat{H}))\,\frac{1}{2}\,\nabla_1\widehat{H}^2\bigr)
\;.
\label{eq-BBC3}
\\
&
\;=\;
-\,\frac{1}{2}\;\widehat{\Tt}\bigl( \widehat{H}\,( f_{\beta}(\widehat{H})-f_{\infty}(\widehat{H}))\,\nabla_1\widehat{H}\bigr)
\;.
\label{eq-BBC4}
\end{align}
Clearly, neither of the terms on the r.h.s. of \eqref{eq-BBC2} is traceclass w.r.t. $\widehat{\Tt}$, so that the passage to \eqref{eq-BBC3} is formal. However, even in \eqref{eq-BBC3} the operator difference $f_{\beta}(\widehat{H})-f_{\infty}(\widehat{H})$ is not traceclass w.r.t. $\widehat{\Tt}$ as the function $E\mapsto f_{\beta}(E)-f_{\infty}(E)$ is not supported in a bulk gap (for the electric current, this difficulty did not arise as one could take $\beta=\infty$). On the other hand, this function is exponentially small for $E$ outside and $\mu$ inside of a bulk gap (for low temperatures). Hence, for an analysis of the low temperature behavior, it is reasonable to introduce a cut-off, restricting the support of the function to a bulk gap. Taking also into account the factor $\widehat{H}$ in \eqref{eq-BBC4}, we therefore set
$$
g_{\beta}(E)\;=\;E(f_{\beta}(E)-f_{\infty}(E))\rho(E)
\;,
$$
where $E\mapsto \rho(E)\in [0,1]$ is a smooth even function supported in the bulk gap and which is equal to $1$ in some open interval containing $E=0$. Neglecting the terms which are exponentially small at low temperature, the thermal edge current density is now
\begin{equation}
\label{eq-ThermEdgeDef}
\hat{\jmath}_H(\beta)
\;=\;
-\,\frac{1}{2}\;\widehat{\Tt}\bigl(g_{\beta}(\widehat{H})\,\nabla_1\widehat{H}\bigr)
\;.
\end{equation}
This formula is  mathematically well-defined as $g_{\beta}(\widehat{H})$ is traceclass w.r.t. $\widehat{\Tt}$ \cite{PS}. Furthermore, one has $g_{\beta}(E)=g_{\beta}(-E)\geq 0$ and
$$
\int_\RM g_{\beta}(E)\,dE
\;=\;
2\,\int_0^\infty g_{\beta}(E)\,dE
\;\approx\;
2 \,\int_0^\infty E\,f_{\beta}(E)\,dE 
\;=\;
\frac{2}{\beta^2}\int^\infty_0 \frac{x\,dx}{e^x+1}
\;=\;
\frac{\pi^2}{6}\,T^2
\;,
$$
where the exponentially small terms were neglected. Using this normalization factor to normalize the function $g_{\beta}$, it follows from Theorem~\ref{theo-edgequant} that 
$$
\hat{\jmath}_H(\beta)
\;\approx\;
\frac{\pi}{24}\,T^2
\;\Ch(P)
\;,
$$
up to corrections which are exponentially small.  The thermal edge conductance $\hat{\kappa}(\beta)$ can now be defined as the  temperature derivative of the thermal boundary currents:
\begin{equation}
\label{eq-ThermalEdgeQuant}
\hat{\kappa}(\beta)
\;=\;
\frac{\pi}{12}\,T
\;\Ch(P)
\;.
\end{equation}
Compairing with \eqref{eq-ThermalEdgeQuant0}, this equation establishes the BBC for the thermal currents. 

\section{Thermo-electric Hall effect}
\label{sec-TEQHE}

For sake of completeness, let us also mention the thermo-electric Ettingshausen-Nernst effect even though there is no non-trivial quantum Hall regime for it. One imposes an external electric field (as in Section~\ref{sec-ChargeCons}) and then measures the transverse thermal current (as in Section~\ref{sec-THE}). By the Onsager relations, one can alternatively look at electric currents as a reaction to a thermal gradient, which is called the Seebeck Hall effect. Again these effects require charge conservation, which is hence imposed here. The derivation of the Kubo formula for perturbation $\Pp=\lambda X_1Q$ and measured current  $\frac{1}{2}\nabla_2 H^2$ is somewhat delicate, just as for the thermal Hall conductance. The result for the linear response coefficient is \cite{SSt}:
$$
\alpha(\beta)
\;=\;
\frac{1}{2}\;\beta\;\int_\RM dE\,E\,f'_\beta(E)
\;\imath\,\Tt
\big(P_E[\nabla_1P_E,\nabla_2 P_E]\big)
\;,
$$
Again by Theorem~\ref{theo-MatterHallKuboChern}, $\Tt\big(P_E[\nabla_1P_E,\nabla_2 P_E]\big)$ is locally constant and therefore \eqref{eq-Sommerfeld} implies that $\alpha(\beta)$ vanishes up to terms of order $T^3$. Actually it vanishes to all orders as $f'_\beta$ is an odd function. To derive a formula for the corresponding boundary currents,  one proceeds as in Section~\ref{sec-ChargeCons} imposing two different chemical potentials $\mu_\pm=\mu\pm\delta$ on the two boundaries of the sample. At zero temperature, one is thus led to consider
$$
\widehat{\Tt}
\Big(
(\hat{p}_--\hat{p}_+)\;\frac{1}{2}\,\nabla_1(\hat{h}-\mu)^2
\Big)
\;=\;
\widehat{\Tt}
\Big(
(\hat{p}_--\hat{p}_+)\,(\hat{h}-\mu)\,\nabla_1\hat{h}
\Big)
\;.
$$
This latter expression vanishes because $E\mapsto \chi_{[\mu-\delta,\mu+\delta]}(E)(E-\mu)$ has vanishing integral. This again establishes a BBC, albeit for vanishing quantities.

\appendix

\section{Conservation equations and currents}
\label{app-Currents}

The object of this appendix is to derive the current operators corresponding to conserved quantities of the Hamiltonian, with a focus on matter, charge, heat and spin current. The basic idea is to transpose the well-known strategy to establish a continuity equation. For thermal and thermoelectric effects, this is, for example, carried out for continuum models in \cite{SSt}. Moreover, continuum BdG Hamiltonians are considered in \cite{VMFT}. This appendix resulted from the apparent lack of a corresponding treatment for lattice operators.

\subsection{Local  densities}

Let us consider one-particle discrete models on $\ell^2(\ZM^d)\otimes\CM^L\otimes\CM^2_{\ph}$ where the fiber $\CM^L=\CM^{2s+1}$ describes a spin $s$ particle and the fiber $\CM^2_{\ph}$ take into account the particle-hole symmetry (for sake of notational simplicity, no further internal degrees of freedom are included). As a basis of the Hilbert space, we choose the common eigenbasis  $|n,l, \eta\rangle$ of the position, spin and charge operators just as in Sections~\ref{sec-BdGgen} and \ref{sec-alg} , namely
$$
X\;|n,l, \eta\rangle\;=\; n\; |n,l, \eta\rangle\;,
\qquad
S^3\;|n,l, \eta\rangle\;=\; l\,\eta\; |n,l, \eta\rangle\;,
\qquad
Q\;|n,l, \eta\rangle\;=\; \eta\; |n,l, \eta\rangle\;,
$$
for $ n\in\ZM^d$, $l\in\{-s,-s+1,\ldots, s-1,s\}$ and $\eta\in\{-1,1\}$. One can expand any operator  $A$ on this basis:
$$
A\;=\;\sum_{n,n'\in\ZM^d}\; \sum_{l,l'=-s}^s\; \sum_{\eta,\eta'=\pm1}\;A_{l,l'}^{\eta,\eta'}(n,n')
\;|n,l,\eta\rangle\langle n',l',\eta'|\;,\qquad\quad A_{l,l'}^{\eta,\eta'}(n,n')\in\CM\;.
$$
It is convenient to introduce the set of matrices $A(n,n')=(A_{l,l'}^{\eta,\eta'}(n,n'))\in{\rm Mat}(\CM^{2L})$ for all $n,n'\in\ZM^d$. With this notation, the selfadjointness of $A$ is equivalent to $A(n,n')=A(n',n)^*$ for all $n,n'\in\ZM^d$. In the following we will use often the short notation
$$
|n\rangle\;A(n,n')\;\langle n'|\;=\;  \sum_{l,l'=-s}^s\; \sum_{\eta,\eta'=\pm1}\;A_{l,l'}^{\eta,\eta'}(n,n')
\;|n,l,\eta\rangle\langle n',l',\eta'|\;.
$$
We need  to express any such operator in terms of \emph{local} quantities. For this purpose let us introduce the on-site projection
\begin{equation}\label{eq:loc_dens1}
\rho(n)\;=\;|n\rangle\langle n|
\;=\; \sum_{l=-s}^s\; \sum_{\eta=\pm1}\;|n,l,\eta\rangle\langle n,l,\eta|
\;.
\end{equation}
%

\begin{defini}[Local density]
Given a bounded  operator $A$ on $\ell^2(\ZM^d)\otimes\CM^L\otimes\CM^2_\ph$ one defines its related \emph{local density} by
\begin{equation}\label{eq:loc_op}
\rho_A(n)\;=\;\frac{1}{2}\;\{A,\rho(n)\}\;=\;\frac{1}{2}\sum_{n'\in\ZM^d}\;\Big[|n\rangle\;A(n,n')\;\langle n'|+|n'\rangle\;A(n',n)\;\langle n|\Big]\;, \qquad\quad n\in\ZM^d\;.
\end{equation}
Therefore the local density $\rho_A$ is an  operator-valued map from $\ZM^d$ into the bounded operators on $\ell^2(\ZM^d)\otimes\CM^L\otimes\CM^2_\ph$.
\end{defini}

The on-site projection \eqref{eq:loc_dens1} is hence the local density of the identity operator
$$
\rho(n)\;=\;\frac{1}{2}\;\{\one,\rho(n)\}\;.
$$
For this reason we refer to $\rho$ as the {local matter density}.  Next the {local charge density} is 
$$
\rho_{\rm Q}(n)\;=\;\frac{1}{2}\;\{Q ,\rho(n)\}\;=\;Q \rho(n)\;=\; \rho(n)Q
\;.
$$
The {local spin density} for a BdG operator is defined by
$$
\rho_{S^3}(n)\;=\;\frac{1}{2}\;\{S^3 ,\rho(n)\}\;=\;S^3 \rho(n)\;=\;\rho(n)S^3  
\;.
$$
Finally, given a (selfadjoint bounded) Hamiltonian $H=\sum_{n,n'\in\ZM^d}\; |n\rangle\; H(n,n')\;\langle n'|$ which induces the dynamics of the system, one defines  the associated {local energy density}
$$
\rho_H(n)\;=\;\frac{1}{2}\;\{H,\rho(n)\}\;=\;\frac{1}{2}\sum_{n'\in\ZM^d}\;\Big[|n\rangle\; H(n,n')\;\langle n'|+|n'\rangle\;H(n',n)\;\langle n|\Big]\;.
$$

\subsection{The gradient operator}

Given a bounded operator $A$ with density $\rho_A(n)$ as in \eqref{eq:loc_op}, the $j$-th  discrete derivative of  $\rho_A$ is defined by
\begin{equation}\label{eqD_i}
(\partial_j\rho_A)(n)\;=\;-\;\frac{ 2}{d}\sum_{n'\in\ZM^d\setminus\{n\}}\;\frac{1}{(n-n')_j}\Big[|n\rangle\;A(n,n')\;\langle n'|+|n'\rangle\;A(n',n)\;\langle n|\Big]\;.
\end{equation}
We refer to the vector $\partial \rho_A=(\partial_1 \rho_A,\ldots,\partial_d \rho_A)$ as the {gradient} of the density $\rho_A$. Given a vector of densities $\rho=(\rho_1,\ldots,\rho_d)$
one then introduces  the \emph{divergence} of $\rho$ by
$$
(\partial \cdot \rho)(n)\;=\; \sum_{j=1}^d (\partial_j\rho_j)(n)\;.
$$

\subsection{Continuity equation for matter}
\label{app-MatterCur}

The {velocity operator} 
${V}=({V}_1,\ldots,{V}_d)$ (with respect to $H$) is given by Heisenberg's equation:
$$
{ V}\;=\;\imath\;[H,X]\;=\;- \imath\,\sum_{n,n'\in \ZM^d}(n-n')_j\, |n\rangle\;H(n,n')\;\langle n'|
\;.
$$
It is  useful to write the components of the velocity as
$$
\begin{aligned}
{ V}_j&\;=\;\sum_{n,n'\in \ZM^d} |n\rangle\; V_j(n,n')\;\langle n'|\;,\qquad\quad V_j(n,n')\;=\;-\imath\,(n-n')_j\; H(n,n')\;.
\end{aligned}
$$
The {density of matter current} $\Jj(n)=(\Jj_1(n),\ldots,\Jj_d(n))$
is, by definition, the local density of the velocity, hence 
\begin{equation}\label{eq:free_cur}
\Jj_j(n)
\;=\;\frac{1}{2}\{{ V}_j,\rho(n)\}
\;=\;-\frac{\imath}{2}\sum_{n'\in \ZM^d}(n-n')_j\,\Big[|n\rangle\;H(n,n')\;\langle n'|-|n'\rangle\;H(n',n)\;\langle n|\Big]\;.
\end{equation}
On the other hand, the time derivative  of the {matter density} $\rho(n)$ is again given by Heisenberg's equation:
$$
\begin{aligned}
\partial_t\,\rho(n)\;&=\;-\imath\;[\rho(n),H]\\
&=\;-\imath \sum_{n'\in\ZM^d}\Big[|n\rangle\;H(n,n')\;\langle n'|-|n'\rangle\;H(n',n)\;\langle n|\Big]\\
&=\;-\imath \sum_{n'\in\ZM^d\setminus\{n\}}\Big[|n\rangle\;H(n,n')\;\langle n'|-|n'\rangle\;H(n',n)\;\langle n|\Big]\;.
\end{aligned}
$$
One has
$$
\sum_{n\in\ZM^d}\partial_t\,\rho(n)\;=\;-\imath\;[\one,H]\;=\; 0
\;,
$$ 
which can be interpreted as the{global conservation of matter} in the system. 
A comparison between \eqref{eqD_i} and \eqref{eq:free_cur} shows that
$$
(\partial_j\Jj_j)(n)\;=\;\frac{\imath}{d}\sum_{n'\in\ZM^d}\Big[|n\rangle\;H(n,n')\;\langle n'|-|n'\rangle\;H(n',n)\;\langle n|\Big]\;,\qquad\forall\; j=1,\ldots,d
$$
and this proves the validity of the  {continuity equation for matter} 
\begin{equation}\label{eq:continuity}
\partial_t\,\rho(n)\;=\;-\,\partial\cdot\Jj(n)
\;.
\end{equation}
Hence the {macroscopic matter current} simply agrees with the velocity
$$
\Jj\;=\;\sum_{n\in\ZM^d}\Jj(n)\;=\;V\;=\;\imath\;[H,X]\;.
$$

\subsection{Continuity equation for charge}

The time derivative  of the {charge density} $\rho_Q(n)$ is given by
$$
\partial_t\,\rho_Q(n)\;=\;-\imath\;[\rho_Q(n),H]
\;.
$$
In order to derive a continuity equation for the charge, one needs to require the {global conservation of charge}, namely  
$$
\sum_{n\in\ZM^d}\partial_t\,\rho_Q(n)\;=\;-\imath\;[Q,H]\;=\; 0\;.
$$ 
This is a constraint on the Hamiltonian which is fulfilled exactly when $H$ is diagonal w.r.t. the grading $\CM^2_\ph$. Using the relation $\rho_Q(n)=Q\rho(n)=\rho(n)Q$ and the fact that $Q$ commutes with the gradient, one immediately obtains form \eqref{eq:continuity} that
$$
\partial_t\,\rho_Q(n)\;=\;Q\partial_t\,\rho(n)\;=\;-Q\;\partial\cdot \Jj(n)\;=\;-\partial\cdot (Q\Jj)(n)\;.
$$
This provides the continuity equation for the {charge density}
\begin{equation}\label{eq:continuity_charge}
\partial_t\,\rho_Q(n)\;=\;-\partial\cdot {\Jj}_Q(n)
\;,
\end{equation}
where the {density  current of charge} is defined by
$$
 {\Jj}_Q(n)\;=\; Q\; \Jj(n)\;=\; \Jj(n)\; Q\;.
$$
Let us observe that the fact of $\Jj(n)$ commuting with the charge operator $Q$ is a consequence of $[X,Q]=[H,Q]=[\rho(n),Q]=0$.  The {macroscopic charge current} is then 
$$
\Jj_Q\;=\;\sum_{n\in\ZM^d}{ \Jj}_Q(n)\;=\;\Jj\;Q\;=\;\imath\;[H,X]\;Q\;.
$$
Due to the above-mentioned commutation relations, one has  $[\Jj_Q,Q]=0$.

\subsection{Continuity equation for spin}
\label{app-SpinCurr}

A continuity equation for the spin {spin density} $\rho_{S^3}(n)$ can be derived by the same strategy. First of all, the time derivative  of $\rho_{S^3}$  is 
$$
\partial_t\,\rho_{S^3}(n)\;=\;-\imath\;[\rho_{S^3}(n),H]
\;.
$$
As a matter of fact, a continuity equation for spin  needs  the {global conservation of spin}:
$$
\sum_{n\in\ZM^d}\partial_t\,\rho_{S^3}(n)\;=\;-\imath\;[S^3,H]\;=\; 0\;.
$$ 
This equation is exactly the case $j=3$ of \eqref{eq-SU2inv} and corresponds to the U$(1)$ invariance of the Hamiltonian. 
Using the relation $\rho_{S^3}(n)=S^3\rho(n)=\rho(n)S^3$ and the fact that $S^3$ commutes with the gradient, one immediately obtains form \eqref{eq:continuity} a continuity equation from the {spin density}
\begin{equation}\label{eq:continuity_spin}
\partial_t\,\rho_{S^3}(n)\;=\;-\partial\cdot {\Jj}_{S^3}(n)
\end{equation}
where the {density  current of spin} is given by
$$
 {\Jj}_{S^3}(n)\;=\; S^3\; \Jj(n)\;=\; \Jj(n)\; S^3\;.
$$
Again it is crucial for this derivation that $[X,S^3]=[H,S^3]=[\rho(n),S^3]=0$. The {macroscopic spin current} is  given by
$$
\Jj_{S^3}\;=\;\sum_{n\in\ZM^d}{ \Jj}_{S^3}(n)\;=\;V\;S^3\;=\;\imath\;[H,X]\;S^3\;
$$
and one can check that $[\Jj_{S^3},S^3]=0$. 

\subsection{Continuity equation for energy}
\label{sec-EnergyCur}

Let $H$ be a Hamiltonian  and $\rho_H(n)$ the associated {energy density}. The time derivative of 
$\rho_H(n)$ via the Heisenberg's equation reads 
 \begin{equation}\label{eq:dens_energ_01}
\partial_t\, \rho_H(n)\;=\;-\imath\;[\rho_H(n),H]\;=\;-\imath\;[\rho(n),\frac{1}{2}H^2]\;.
\end{equation}
Since 
$$
\sum_{n\in\ZM^d}\partial_t\,\rho_H(n)\;=\;-\imath\;[H,H]\;=\; 0
$$ 
we have a
{global conservation of energy} and we can proceed to the derivation of a continuity equation for the energy density. Equation \eqref{eq:dens_energ_01} can be seen as the time derivative of the matter density $\rho(n)$ with respect to the ``effective'' Hamiltonian
$$
H'\;=\;\frac{1}{2}\;H^2\;=\;\sum_{n,n'\in\ZM^d}\;  |n\rangle\;H'(n,n')\;\langle n'|\;,\qquad\quad 
H'(n,n')\;=\; \frac{1}{2}\sum_{m\in\ZM^d}H(n,m) H(m,n')\;.
$$
Then we can immediately use the continuity equation
\eqref{eq:continuity} for $H'$  which now reads
\begin{equation}\label{eq:continuity_energy_02}
\partial_t\,\rho_H(n)\;=\;-\partial\cdot \Jj_H (n)
\end{equation}
where now the current density is defined by the Hamiltonian $H'$, namely
\begin{equation}
\Jj_H (n)\;=\;-\frac{\imath}{2}\Big\{[X,H'],\rho(n)\Big\}\;=\;-\frac{\imath}{4}\Big\{[X,H^2],\rho(n)\Big\}
\end{equation}
We refer to $\Jj_H(n)$ as the {density of current of energy}. The {macroscopic energy current} is by definition
$$
\Jj_H\;=\;\sum_{n\in\ZM^d}\Jj_H(n)\;=\;\frac{\imath}{2}\;[H^2,X]\;.
$$

\section{Kubo formula}
\label{app-kubo}

Let $H=H^*$ be some bounded one-particle BdG Hamiltonian on $\ell^2(\ZM^2,\CM^{2L})$ describing a system of independent fermions. Even if there is not particle-hole creation so that $H$ is diagonal and one is in Class A, we will pass to the BdG representation and absorb the chemical potential $\mu$ in the Hamiltonian as in Section~\ref{sec-BdGgen}. We voluntarily skip mathematical details in the following, which can readily filled in for a given special case ({\it e.g.} as in \cite{SB,DL}).  The Hamiltonian generates a time evolution of density matrices $f(t)$ according to the Liouville equation 
$$
\partial_t f(t)
\;=\;
\Ll_H(f(t))
\;,
\qquad
\Ll_H(A)\;=\;\imath\,[A,H]\;.
$$
Here $\Ll_H$ is called the Liouville operator. Note that this evolution leaves invariant the algebra generated by $H$ and therefore, in particular, the equilibrium distribution $f_{\beta}(H)$ given by the Fermi-Dirac function $f_{\beta}(E)=(1+e^{\beta E})^{-1}$. Now let $\lambda\,\Pp$ be a possibly unbounded perturbation of $H$, where $\lambda\geq 0$ is a coupling constant. This changes the time evolution of density matrices to 
$$
\partial_t f(t)
\;=\; \Ll_{H+\lambda\Pp}(f(t))
\;=\;
(\Ll_H+\lambda\,\Ll_\Pp)(f(t))
\;.
$$
The perturbation has to be such that $\Ll_H+\lambda\,\Ll_\Pp$ is indeed the generator of time-evolution (automorphism of the operator algebra). Then the solution can simply be written as
$$
f(t)
\;=\;
e^{t(\Ll_H+\lambda\,\Ll_\Pp)} (f(0))
\;.
$$
As initial condition, let us choose $f(0)=f_{\beta}(H)$.  Now one wants to measure the current associated to a (bounded) observable $\Jj=\Jj^*$. Furthermore, an exponential time average of time scales of order $\frac{1}{\delta}$ will be taken. If $\Tt$ is an adequate trace (as the one in Section~\ref{sec-trace}) and one takes into account the factor $\frac{1}{2}$ of \eqref{eq-Fac1/2}, then this time-averaged current is given by
$$
j(\lambda,\delta)
\; = \;
\lim_{\delta\downarrow 0}
\;\delta\;\int_0^\infty dt\;e^{-\delta\,t}\;
\frac{1}{2}\;\Tt\bigl(f(t)\,\Jj \bigr)
\;=\;
\lim_{\delta\downarrow 0}\;
\frac{\delta}{2}\;\Tt\bigl(
(\delta-\Ll_{H}-\lambda\,\Ll_\Pp)^{-1}(f_{\beta}(H))
\;
\Jj\bigr)
\;.
$$
If the limit $\delta\to 0$ exists, one then sets
$$
j(\lambda)
\; = \;
\lim_{\delta\downarrow 0}
j(\lambda,\delta)
$$
However, this limit may not always exist and, if it does not, $\delta$ effectively introduces some dissipation into the system and one can see $\frac{1}{\delta}$ as the associated relaxation time (resulting from inelastic scattering processes \cite{SB}). For dissipationless currents such as the Hall conductance and Hall spin conductance, one can then show that the limit indeed does exist.  Let us now further calculate $j(\lambda,\delta)$. If $\lambda=0$ and $f(0)=f_{\beta}(H)$ as above, then
$$
j(0,\delta)
\;=\;
\frac{1}{2}\;
\Tt\bigl(f_{\beta}(H)\,\Jj \bigr)
\;=\;
\frac{1}{2}\;
\delta\;
\Tt\bigl(
(\delta-\Ll_{H})^{-1}(f_{\beta}(H))
\;
\Jj\bigr)
\;.
$$
Let us assume from now that this current at equilibrium vanishes. Hence subtracting this vanishing term $j(0)=0$, one finds from the resolvent formula
\begin{eqnarray*}
j(\lambda,\delta)
& = &
\frac{1}{2}\;
\lambda\,\Tt\bigl(
(\delta-\Ll_{H}-\lambda\,\Ll_\Pp)^{-1}(\Ll_\Pp(f_{\beta}(H)))
\;
\Jj\bigr)
\\
& = &
\frac{1}{2}\;
\lambda\,\Tt\bigl(
\Ll_\Pp(f_{\beta}(H))
\;
(\delta+\Ll_{H}+\lambda\,\Ll_\Pp)^{-1}(\Jj)\bigr)
\;,
\end{eqnarray*}
where in the second equality it was used that $\Ll_H$ and $\Ll_\Pp$ are anti-selfadjoint as superoperators on $L^2(\Aa,\Tt)$ and it was used that $\Tt$ is invariant under $\Ll_\Pp$. The linear response coefficient $\sigma_{\Pp,\Jj}(\beta)$ is defined by
$$
j(\lambda,\delta)
\;=\;
\lambda\,\sigma_{\Pp,\Jj}(\beta,\delta)
\;+\;
\Oo(\lambda^2)
\;.
$$
Due to the above one deduces the Kubo formula
\begin{equation}
\label{eq-Kubogen}
\sigma_{\Pp,\Jj}(\beta,\delta)
\; = \;
\frac{1}{2}\;
\Tt\bigl(
\Ll_\Pp(f_{\beta}(H)) 
\;(\delta+\Ll_H)^{-1}(\Jj) \bigr)
\;.
\end{equation}
Let us consider the case where $\Jj=\Ll_{\Pp'}(H)$. Then one has the so-called Onsager relation
$$
\sigma_{\Pp,\,\Ll_{\Pp'}(H)}(\beta,\delta)
\;=\;
-\;
\sigma_{\Pp',\,\Ll_{\Pp}(H)}(\beta,\delta)
\;,
$$
provided the limit in \eqref{eq-Kubogen} exists. Indeed, because $H$ is bounded, one can write $f_{\beta}(H)=g(H)$ for some function in the Schwartz space which has a Fourier transform $\widehat{g}$ such that 
$$
f_{\beta}(H)\;=\;\int dt \;\widehat{g}(t)\,e^{\imath tH}
\;.
$$
Using Duhamel's formula it follows
\begin{eqnarray*}
\sigma_{\Pp,\,\Ll_{\Pp'}(H)}(\beta,\delta)
& = &
\frac{1}{2} \int dt \;\widehat{g}(t)\,
\int^t_0ds\;
\Tt\bigl(
e^{\imath (t-s)H}\,\imath\,
\Ll_\Pp(H)\, e^{\imath sH}\,
\;(\delta+\Ll_H)^{-1}(\Ll_{\Pp'}(H)) \bigr)
\\
& = &
\frac{1}{2}\int dt \;\widehat{g}(t)\,
\int^t_0ds\;
\Tt\bigl(
(\delta-\Ll_H)^{-1}((\Ll_\Pp(H))\; e^{\imath sH}\,\imath\,\Ll_{\Pp'}(H)\,e^{\imath (t-s)H}
 \bigr)
\;,
\end{eqnarray*}
where in the second equality it was used that $\Ll_H$ commutes with functions of $H$ and is anti-self-adjoint on $L^2(\Aa,\Tt)$. Recombining and using the cyclicity of the trace one obtains the Onsager relation.

\section{Quadratic commutator identity}
\label{app-commutator}

Let $\Hh$ be a separable complex Hilbert space with orthonormal basis $(|n\rangle)_{n\geq 1}$. The associated creation and annihilation operator on the fermionic Fock space $\Ff$ are denoted by $\frakc_n^*$ and $\frakc_n$. Let us consider a quadratic operator in creation and annihilation operators of the form
$$
{\bf A}
\;=\;
\frac{1}{2}\;\begin{pmatrix} \frakc^* & \frakc\end{pmatrix}
A
\begin{pmatrix} \frakc \\ \frakc^*\end{pmatrix}
\;,
\qquad
A\;=\;\begin{pmatrix}  \alpha & \beta \\ \gamma & -\alpha^T \end{pmatrix}
\;,
$$
where $\alpha$, $\beta$ and $\gamma$ are operators on $\Hh$.  Here the same notations as in Section~\ref{sec-BdGfirstquant} are used. 
One may assume that  $\beta$ and $\gamma$ are anti-symmetric because the symmetric components $\beta+\beta^T$ and $\gamma+\gamma^T$ lead to vanishing contributions due to the CAR relations. Now let ${\bf A'}$ be a second operator of the same form as ${\bf A}$ with coefficient matrices $\alpha'$, $\beta'$ and $\gamma'$ giving $A'$. Now the commutator $[{\bf A},{\bf A'}]$ turns out to be again quadratic in the creation and annihilation operators which can be expressed in terms of
$$
[A,A']
\;=\;
\begin{pmatrix}  
[\alpha,\alpha']+(\beta\gamma'-\beta'\gamma) & \alpha\beta'+\beta'\alpha^T-\alpha'\beta-\beta(\alpha')^T \\ \gamma\alpha'+(\alpha')^T\gamma-\gamma'\alpha-\alpha^T\gamma' & -[\alpha,\alpha']^T-(\beta\gamma'-\beta'\gamma)^T
\end{pmatrix}
\;.
$$
More precisely, the following well-known fact holds, {\it e.g.} \cite{HHZ}.
%

\begin{proposi}
\label{prop-LieAlg}
The map $A\mapsto {\bf A}$ is a Lie algebra homomorphism, namely
\begin{equation}
\label{eq-commutatorid}
[{\bf A},{\bf A'}]
\,=\,
\frac{1}{2}\,
\begin{pmatrix} \frakc^* & \frakc\end{pmatrix}
[A,A']
\begin{pmatrix} \frakc \\ \frakc^*\end{pmatrix}
\;.
\end{equation}
\end{proposi}

\noindent {\bf Proof.}
The calculatory proof of this identity is just a matter of patience. One way to proceed is to use the linearity of the commutator and consider cases with ${\bf A}$ and ${\bf A'}$ with only one entry each. For example, if $\alpha=\alpha'=\gamma=\beta'=0$, then $4[{\bf A},{\bf A'}]=[\frakc^*\beta \frakc^*,\frakc\gamma' \frakc]$ and writing out gives, with summation over all indices,
$$
4[{\bf A},{\bf A'}]
\;=\;
\sum \beta_{n,n'}\gamma'_{m,m'}(\frakc_n^*\frakc_{n'}^*\frakc_m \frakc_{m'}-\frakc_m \frakc_{m'}\frakc_n^* \frakc_{n'}^*)
\;.
$$
Now using the commutation relations twice in order to move the annihilation operators to the left and right shows that the forth order terms cancel and
$$
4[{\bf A},{\bf A'}]
\,=\,
\sum \beta_{n,n'}\gamma'_{m,m'}(\delta_{n',m}\frakc_n^*\frakc_{m'}  -\delta_{n,m} \frakc_{n'}^* \frakc_{m'}-\delta_{n,m'} \frakc_m \frakc_{n'}^* + \delta_{m',n'} \frakc_m \frakc_{n}^*)
\,=\,
2\,\frakc^*\beta\gamma' \frakc-2\,\frakc \beta\gamma'  \frakc^*
\,.
$$
This indeed gives the corresponding contributions in \eqref{eq-commutatorid}. The other terms can be treated similarly. For the off-diagonal terms, one also has to use the identities $\frakc(\alpha')^T\gamma \frakc=-\frakc\gamma^T\alpha' \frakc=\frakc\gamma \alpha' \frakc$.
\hfill $\Box$

\section{Finite dimensional representations of $\mbox{\rm SU}(2)$}
\label{app-su2}

The Lie algebra  $\mbox{\rm su}(2)$ has three generators $s^1,s^2, s^3$ satisfying $[s^1,s^2]=\imath\,s^3$, $[s^2,s^3]=\imath\,s^1$ and $[s^3,s^1]=\imath\,s^2$. The algebra $\mbox{\rm su}(2)$ has irreducible representations in any dimension $L=2s+1$ for all $s\in\NM/2$ which are then called spin $s$ representation. The concrete spin $s$ representation used here (again denoted by the same letters as the abstract $\mbox{\rm su}(2)$ generators) can be expressed in terms of the $L\times L$ real matrices 
$$
s^3
\;=\;
\left(\begin{array}{ccccc}
 s& 0&\ldots&0&0\\
  0&s-1&\ldots& 0& 0\\
 \vdots&\vdots&\ddots&\vdots & \vdots\\
0&0&\ldots& 1-s& 0\\ 
 0&0&\ldots& 0& -s
 \end{array}\right)
\;,
\qquad
s_+
\;=\;
\left(\begin{array}{ccccc}
 0& \alpha_1&\ldots&0&0\\
  0&0&\alpha_2& \ldots& 0\\
 \vdots&\vdots&\ddots&\ddots & \vdots\\
0&0&\ldots& 0& \alpha_{2s}\\ 
 0&0&\ldots& 0& 0
 \end{array}\right)
 \;,
$$
where
$$
\alpha_{l}\;=\;\sqrt{l(2s+1-l)}\;,
\qquad l=1,\ldots,2s\;.
$$
Then one sets 
$$
s_-\;=\;(s_+)^*\;,
\qquad
s^1\;=\;\frac{1}{2}(s_-+s_+)\;,
\qquad
s^2\;=\;\frac{\imath}{2}(s_--s_+)
\;.
$$
It is matter of direct calculation to verify
$$
[s^3,s_+]\;=\;s_+\;,
\qquad[s^3,s_-]\;=\;-s_-\;,
\qquad[s_+,s_-]\;=\;2\,s^3
\;,
$$
which imply that indeed the $\mbox{\rm su}(2)$ relations are satisfied.  Furthermore the following properties hold for $j=1,2,3$:
$$
(s^j)^*\;=\;s^j\;,
\qquad 
\overline{s^j}\;=\;(-1)^{1+j}\ s^j\;,
\qquad \Tr(s^j)\;=\;0
\;.
$$
Note that for $s=\frac{1}{2}$ the above representation is given by the Pauli matrices $s^1=\frac{1}{2}\left(\begin{smallmatrix} 0 & 1 \\ 1 & 0 \end{smallmatrix}\right)$, $s^2=\frac{1}{2}\left(\begin{smallmatrix} 0 & -\imath \\ \imath & 0 \end{smallmatrix}\right)$ and $s^3=\frac{1}{2}\left(\begin{smallmatrix} 1 & 0 \\ 0 & -1 \end{smallmatrix}\right)$



\begin{thebibliography}{99}
\bibliographystyle{unsrt}








\bibitem[AZ]{AZ} A.~Altland, M.~Zirnbauer, {\sl Non-standard symmetry classes in mesoscopic normal-superconducting hybrid structures}, Phys. Rev. {\bf B 55}, 1142-1161 (1997).


\bibitem[ASS]{ASS} J.~E.~Avron, R.~Seiler, B.~Simon, {\sl Charge Deficiency, Charge Transport and Comparison of Dimensions},
Commun. Math. Phys. {\bf 159}, 399-422 (1994).

\bibitem[BLS]{BLS} V.~Bach, E.~H.~Lieb, J.~P.~Solovej, {\sl Generalized Hartree-Fock theory and the Hubbard model}, J. Stat. Phys. {\bf 76}, 3-89 (1996).

\bibitem[Bel]{Bel} J. Bellissard, {\sl K-theory of C$^*$-algebras in solid state physics}, in {\sl Statistical Mechanics and Field Theory: Mathematical Aspects}, {\sl Lecture Notes in Physics} {\bf 257}, T.~Dorlas, M.~Hugenholtz, M.~Winnink (Eds.), 99-156 (Springer, Berlin, 1986); see also {\sl Coherent and dissipative transport in aperiodic solids},  Lecture Notes in Physics {\bf 597}, P.~Garbaczewski, R.~Olkiewicz (Eds.), 413-486 (Springer, Berlin, 2003).



\bibitem[BES]{BES} J.~Bellissard, A.~van Elst, H.~Schulz-Baldes,
{\sl The Non-Commutative Geometry of the Quantum Hall Effect},
J. Math. Phys. {\bf 35}, 5373-5451 (1994).



\bibitem[BR]{BR} J.~P.~Blaizot, , G. Ripka, {\sl Quantum Theory of Finite Fermi Systems}, (MIT Press, Boston, 1985).


\bibitem[BCR]{BCR} C.~Bourne, A.~L.~Carey, A.~Rennie, {\sl A noncommutative framework for topological insulators}, Rev. Math. Phys. {\bf 28}, 1650004 (2016).




\bibitem[CRK]{CRK} J.~T.~Chalker, N.~Read, V.~Kagalovsky, B.~Horovitz,
Y.~Avishai, A.~W.~W.~Ludwig, {\sl Thermal metal in network models of a disordered two-dimensional superconductor}, Phys. Rev. {\bf B 65}, 012506 (2002).


\bibitem[CHR]{CHR} N.~R.~Cooper, B.~I.~Halperin, I.~M.~Ruzin, {\sl Thermoelectric response of an interacting two-dimensional electron gas in a quantizing magnetic field}, Phys. Rev. {\bf B 55}, 2344 (1997).

\bibitem[DDS]{DDS} G.~De~Nittis, M.~Drabkin, H.~Schulz-Baldes, {\sl Localization and Chern numbers for weakly disordered BdG operators}, Markov Processes Relat. Fields {\bf 21}, 463-482 (2015).

\bibitem[DL]{DL} G.~De~Nittis, M.~Lein, {\sl Linear Response Theory: A Modern Analytic-Algebraic Approach}, (Springer International, Switzerland, 2017).


\bibitem[DS]{DS} G.~De~Nittis, H.~Schulz-Baldes, {\sl Spectral flows associated to flux tubes}, Annales H. Poincare {\bf 17}, 1-35 (2016).

\bibitem[EG]{EG} P. Elbau, G.-M. Graf, {\sl Equality of bulk and edge Hall conductance revisited},  Commun. Math. Phys. {\bf 229}, 415--432 (2002). 

\bibitem[ESS]{ESS} A.~Elgart, M.~Shamis, S.~Sodin, {\sl Localization for non-monotone Schr\"odinger operators}, 
J. European Math. Soc. {\bf 16}, 909-924 (2014).

\bibitem[FT]{FT} M.~Franz, Z.~Tesanovic, {\sl Quasiparticles in the vortex lattice of unconventional superconductors: Bloch waves or Landau levels?}, Phys. Rev. Lett. {\bf  84}, 554-557 (2000).

\bibitem[dG]{dG} P.~de~Gennes, {\sl Superconductivity of metals and alloys}, second edition, (Addison-Wesley, Redwood City, 1989).

\bibitem[GM]{GM} M.~Gebert, P.~M\"uller, {\sl Localization for random block operators}, p. 229-246 in 
{\sl Operator Theory: Advances and Applications} {\bf 232}, (Springer, Berlin, 2013). 


\bibitem[GS]{GS} J.~Grossmann, H.~Schulz-Baldes, {\sl Index pairings in presence of symmetries with applications to topological insulators}, Commun. Math. Phys.  {\bf 343}, 477-513 (2016).

\bibitem[GLR]{GLR} I.~A.~Gruzberg, A.~W.~W.~Ludwig, N.~Read, {\sl Exact exponents for the spin quantum Hall transition}, Phys. Rev. Lett. {\bf 82}, 4524-4527 (1999).


\bibitem[HHZ]{HHZ} P.~Heinzner, A. Huckleberry,  M.~Zirnbauer, {\sl Symmetry classes of disordered fermions}, Commun. Math. Phys.  {\bf 257}, 725-771 (2005).



\bibitem[KRS]{KRS} J.~Kellendonk, T.~Richter, H.~Schulz-Baldes, 
{\sl Edge current channels and Chern numbers in the integer quantum
Hall effect}, Rev. Math. Phys. {\bf 14}, 87-119 (2002).


\bibitem[Kit]{Kit0} A.~Kitaev, {\sl Anyons in an exactly solved model and beyond}, Annals of Physics {\bf 321}, 2-111 (2006).




\bibitem[Lut]{Lut}
J.~M.~Luttinger, {\sl Theory of Thermal Transport Coefficients},
Phys. Rev. {\bf 135}, A1505 (1964)


\bibitem[MT]{MT} V.~Mathai, G.~C.~Thiang, {\sl $T$-Duality Simplifies BulkÐBoundary Correspondence}, Commun. Math. Phys. {\bf 345}, 675-701 (2016).

\bibitem[NRN]{NRN} R.~Nakai, S.~Ryu, K.~Nomura, {\sl Finite-temperature effective boundary theory of the quantized thermal Hall effect}, New Journal of Physics {\bf 18}, 023038 (2016).

\bibitem[OS]{OS} H.~Oji, P.~Streda, {\sl Theory of electronic thermal transport: magnetoquantum corrections to the thermal transport coefficients}, Phys. Rev. {\bf B 31}, 7291-7295 (1985).




\bibitem[PS]{PS}
E.~Prodan, H.~Schulz-Baldes, {\sl Bulk and Boundary Invariants for Complex Topological Insulators: From $K$-Theory to Physics}, (Springer International, Switzerland, 2016).


\bibitem[QNS]{QSN} T.~Qin, Q.~Niu, J.~Shi, {\sl Energy Magnetization and the Thermal Hall Effect}, Phys. Rev. Lett. {\bf 107}, 236601 (2011).

\bibitem[RG]{RG} N.~Read, D.~Green, {\sl Paired states of fermions in two dimensions with breaking of parity and time-reversal symmetries and the fractional quantum Hall effect}, Phys. Rev. {\bf B 6}, 10267-10297 (2000).





\bibitem[Sca]{Sca} D.~J.~Scalapino, {\sl The case for $d_{x^2-y^2}$  pairing in the cuprate superconductors}, Physics Reports {\bf 250},  329-365 (1995).

\bibitem[SRFL]{SRFL}
A.~P.~Schnyder, S.~Ryu, A.~Furusaki, A.~W.~W.~Ludwig,
{\sl Classification of topological insulators and superconductors in three spatial dimensions},
Phys. Rev. {\bf B 78}, 195125-295144 (2008).


\bibitem[SB]{Sch2} H.~Schulz-Baldes, {\sl Topological insulators from the perspective of non-commutative geometry and index theory}, 
Jahresber. Dtsch. Math.-Ver.  {\bf 118}, 247Ð273 (2016). 

\bibitem[SBB]{SB}  H.~Schulz-Baldes, J.~Bellissard, {\sl A Kinetic Theory for Quantum Transport in Aperiodic Media},  
J. Stat. Phys. {\bf 91}, 991-1027 (1998).

\bibitem[SKR]{SKR} H.~Schulz-Baldes, J.~Kellendonk, T.~Richter, {\sl Similtaneous quantization of edge and bulk Hall conductivity},  J. Phys. {\bf A 33}, L27-L32 (2000).

\bibitem[ST]{ST} H.~Schulz-Baldes, S.~Teufel, {\sl Orbital polarization and magnetization for independent particles in disordered media},  Commun. Math. Phys. {\bf 319}, 649-681 (2013).

\bibitem[SMF]{SMF}
T.~Senthil, J.~B.~Marston, M.~P.~A.~Fisher, 
{\sl Spin quantum Hall effect in unconventional superconductors},
Phys. Rev. {\bf B 60}, 4245-4254 (1999).

\bibitem[SF]{SFi}
T.~Senthil, M.~P.~A.~Fisher, 
{\sl Quasiparticle localization in superconductors with spin-orbit scattering}, Phys. Rev. {\bf B 61}, 9690-9699 (2000).


\bibitem[SYN]{SYN} Y.~Shimizu, A.~Yamakage, K.~Nomura, {\sl Quantum thermal Hall effect of Majorana fermions on the surface of superconducting topological insulators}, Phys. Review {\bf B 91}, 195139 (2015).


\bibitem[SSt]{SSt}  L.~Smrcka, P.~Streda, {\sl Transport coefficients in strong magnetic fields}, J. Phys. {\bf C 10}, 2153-2161 (1977).

\bibitem[SuF]{SF} H.~Sumiyoshi, S.~Fujimoto, {\sl Quantum Thermal Hall Effect in a Time-Reversal-Symmetry-Broken Topological Superconductor in Two Dimensions : Approach From Bulk Calculations}, J. Phy. Soc. Jap.  {\bf 82}, 023602 (2013).

\bibitem[Sh]{Sh} D.~Shale, {\sl Linear symmetries of free boson fields}, Trans. Amer. Math. Soc.  {\bf 103}, 149-167 (1962).



\bibitem[Vis]{Vis} A.~Vishwanath, 
{\sl Dirac Nodes and Quantized Thermal Hall Effect in the Mixed State of d-wave Superconductors}, Phys, Rev. {\bf B 66}, 064504 (2002).

\bibitem[VMT]{VMFT} O.~Vafek, A.~Melikyan,  Z.~Te\v{s}anovi\'{c}, {\sl Quasiparticles Hall transport of $d$-wave superconductors in the vortex state}, Phys, Rev. {\bf B 64}, 224508 (2001).


\bibitem[WSS]{WSS} S.~R.~White, D.~J.~Scalapino, R.~L.~Sugar, N.~E.~Bickers, R.~T.~Scalettar, {\sl Attractive and repulsive pairing interaction vertices for the two-dimensional Hubbard model}, Phys. Rev. {\bf B 39}, 839-842 (1989).

\end{thebibliography}
\end{document}